\documentclass[aps,twocolumn,pra,superscriptaddress,amsmath,showpacs,tightenlines,nofootinbib]{revtex4}
\usepackage{epsfig,graphicx,times}
\usepackage{amstext}
\usepackage{amsmath}            
\usepackage{amssymb}            
\usepackage{graphicx}           
\usepackage{latexsym}
\usepackage{bm}
\usepackage{color}
\begin{document}
\title{Energy localization enhanced ground-state cooling of mechanical resonator from room temperature in optomechanics using a gain cavity}
\author{Yu-Long Liu}
\affiliation{Institute of Microelectronics, Tsinghua University, Beijing 100084, China}
\affiliation{Tsinghua National Laboratory for Information Science and Technology (TNList),
Beijing 100084, China}
\author{Yu-xi Liu}
\email{yuxiliu@mail.tsinghua.edu.cn}
\affiliation{Institute of Microelectronics, Tsinghua University, Beijing 100084, China}
\affiliation{Tsinghua National Laboratory for Information Science and Technology (TNList),
Beijing 100084, China}
\date{\today }

\begin{abstract}
When a gain system is coupled to a loss system, the energy usually flows from the gain system to the loss one. We here present a counterintuitive theory for the ground-state cooling of the mechanical resonator in optomechanical system via a gain cavity. The energy flows first from the mechanical resonator into the loss cavity, then into the gain cavity, and finally localizes there. The energy localization in the gain cavity dramatically enhances the cooling rate of the mechanical resonator. Moreover, we show that unconventional optical spring effect, e.g., giant frequency shift and optically induced damping of the mechanical resonator, can be realized. Those feature a pre-cooling free ground-state cooling, i.e., the mechanical resonator in thermal excitation at room temperature can directly be cooled to its ground state. This cooling approach has the potential application for fundamental tests of quantum physics without complicated cryogenic setups.
\end{abstract}

\maketitle
\pagenumbering{arabic}

\section{Introduction}
Mechanical resonator provides an ideal platform to study quantum mechanics of macroscopic objects~\cite{CM1,CM2,CM3}. It can also be used to explore the quantum-classical boundary~\cite{review1}, and applied to high-precision metrology and quantum information processing~\cite{review2}. It is known that the thermal phonon significantly affects the mechanical resonator in low frequency. Thus, it is very important to remove the phonon effect in many applications of quantum physics by cooling the mechanical resonator to its quantum ground state. Various experimental methods have been applied to cool the mechanical resonator, e.g., pure cryogenic cooling~\cite{cryogenic cooling1,cryogenic cooling2,cryogenic cooling3}, electronic feedback cooling~\cite{electronic feedback1,electronic feedback2}, active~\cite{active optical feedback1,active optical feedback2,active optical feedback3,active optical feedback4,active optical feedback5} and passive optical feedback cooling~\cite{passive optical feedback1}, as well as cavity optomechanical backaction cooling, which includes resolved-sideband and unresolved-sideband cooling~\cite{backaction optomechanical review1,bc1,bc2,bc3,bc4,bc5,bc6,bc7,bc8,bc9,bc10,bc11,bc12,bc13,bc14,bc15}.

In past few years, theoretical ~\cite{theoretical1,theoretical2,theoretical3,theoretical4,theoretical5,theoretical6,theoretical7,theoretical8,Liyong2008,Liyong2011}
and experimental~\cite{resolved sideband cooling0,resolved sideband cooling1,resolved sideband cooling2,resolved sideband cooling3} works have shown that the resolved-sideband cooling is a feasible approach toward the quantum ground state of macroscopic mechanical resonators in optomechanics. Here, the resolved-sideband means that the decay rate of the optical mode is much smaller than the frequency of the mechanical resonator. However, such cooling mechanism still puts a cooling limit (minimum phonon number in the steady state) in principle~\cite{coolingreview}. Fundamentally, this is determined by the competition between cooling and heating processes of the mechanical resonator.

For most of standard optomechanical systems in which loss cavities are coupled to mechanical resonators via radiation pressure, the mechanical resonators are always in low frequency (e.g., from kHz to MHz)~\cite{FP}. Thus, the heating process is mainly determined by the thermalization of the mechanical resonator via the interaction with its thermal reservoir. However, the sideband cooling does not sufficiently overwhelm such thermalization, especially when the mechanical resonator is in the room temperature. To achieve the ground-state cooling with the sideband cooling, great efforts have been made to improve the phonon cooling rate or reduce the initial thermal phonon number by putting various optomechanical systems, e.g., microtoroid~\cite{ground state backaction cooling1}, photonic-crystal~\cite{ground state backaction cooling2}, and electromechanical-circuit~\cite{ground state backaction cooling3,ground state backaction cooling4,ground state backaction cooling5}, into the cryogenic environment. Such cryogenic environment is always supported by a continuous-flow helium cryostat or a helium exchange gas cryostat, which imposes severe technical challenges and fundamental constraints. Without the need for cryogenic precooling, one could extend their functions as hybrid quantum systems with atomic ensembles or single atom~\cite{hybridr0,hybridr1,hybridr2,hybridr3}. Hybrid systems would also open up practical avenues for applications of such quantum optomechanical systems, e.g., optically and mechanically linked hybrid quantum networks operating at the room temperature~\cite{qinternet1,qinternet2}. Thus, how to get ground-state cooling of the mechanical resonator without cryogenic precooling is an important subject to be explored.

Essentially, two methods can be used to solve the cooling problem without precooling: (i) greatly reducing the coupling strength between the mechanical mode and the thermal reservoir~\cite{lownoise1,lownoise2}, this requires a ultralow decay rate of the mechanical resonator (e.g.,$\sim$ mHz); (ii) significantly enhancing the phonon cooling rate. However, the first method is difficult to be extended to all types of standard optomechanical systems. Thus, various approaches have been proposed to improve the cooling rate, e.g., cooling in the strong coupling regime~\cite{st1,st2}, hybrid cooling with atomic systems~\cite{hybrid0,hybrid1,hybrid2,hybrid3,hybrid4,hybrid5,hybrid6,hybrid7} or artificial atoms~\cite{Zhang2005,Wang2009,Li2011}, cooling by electromagnetically-induced-transparency~\cite{EIT1,EIT2,EIT3}, squeezed optical fields~\cite{squeeze1,squeeze2,squeeze3}, dark mode interaction~\cite{DM1,DM2}, dissipative optomechanical coupling~\cite{diss1,diss2,diss3,diss4,diss5,diss6,diss7,diss8,diss9,diss10,diss11}, pulsed excitation~\cite{pulsecooling1,pulsecooling2,pulsecooling3,pulsecooling4,pulsecooling5,pulsecooling6}, spontaneous Brillouin scattering~\cite{br1,br2}, cavity polaritons~\cite{polariton1,polariton2,polariton3,polariton4}, and Non-Markovian evolution~\cite{nm1,nm2}. Unfortunately, the phonon cooling rate in these approaches is still not high enough to overcome the heating rate at the room temperature, and the cryogenic precooling operation is still required.

Recently, the field localization effects have been observed in the coupled-optical-cavity system~\cite{scienceBo}, with a $\mathcal{PT}$-symmetric structure~\cite{PT2,PT2s}. It was shown~\cite{PT2} that the field localization is much faster than the dissipation process. Inspired by these studies~\cite{scienceBo,PT2,PT2s}, we present a ground state cooling method for the mechanical resonator in optomechanics using the field localization effect. Here, we introduce a gain cavity, which is coupled to the standard optomechanical system. This approach allows ground-state cooling of the mechanical resonator without cryogenic precooling. Thus, it may make the quantum experiments feasible at the room temperature.

The paper is organized as follows. In Sec.~II, we describe in detail the Hamiltonian and the quantum Langevin equations of the proposed model with $\mathcal{PT}$-symmetry. In Sec.~III, we derive and calculate the spectral density of the optical force. By using the Fermi's golden rule, the cooling rate of the mechanical resonator assisted by one active cavity is also discussed. In Sec.~IV, we study the relation between the phase transition and the energy localization enhanced optomechanical cooling. In Sec.~V, we study the active cavity assisted optical spring effect. In Sec.~VI, we study the cooling limit and show a precooling-free ground-state cooling of the mechanical resonator at the room temperature. Finally, in Sec. VII, we conclude our work.
\begin{figure}[ptb]
\includegraphics[bb=85 159 607 350, width=8cm, clip]{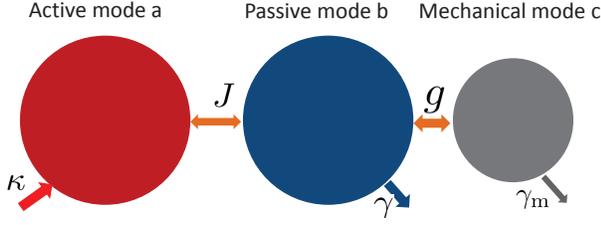}\caption{(Color online) Active-cavity assisted optomechanical system. An active optical mode $a$ with gain $\kappa$ is coupled to a passive optical mode $b$ with decay rate $\gamma$, where $J$ represents the coupling strength. Here, $g$ is the optomechanical coupling strength between the optical mode $b$ and the mechanical mode $c$ with decay a rate $\gamma_{\mathrm{m}}$.}
\label{fig1}
\end{figure}

\section{Model}
As schematically shown in Fig.~\ref{fig1}, we study a system, which consists of two coupled optical cavities and a mechanical resonator. One of the cavities has the gain (called as active cavity), and is described by the bosonic annihilation $a$ and creation $a^{\dag}$ operators. Its frequency and gain are represented by $\omega_{\mathrm{a}}$ and $\kappa$, respectively. The other one is a lossy cavity (called as passive cavity) with the frequency $\omega_{\mathrm{b}}$ and the decay rate $\gamma$, and is described by the bosonic annihilation $b$ and creation $b^{\dag}$ operators. We assume that the coupling strength $J$ between the two cavities can be effectively modulated, e.g., by the distance between them. The mechanical resonator, with the resonance frequency $\omega_{\mathrm{m}}$ and the decay rate $\gamma_{\mathrm{m}}$, is described by bosonic annihilation $c$ and creation $c^{\dag}$ operators. The loss cavity is optomechanically coupled to the mechanical resonator with the single-photon coupling strength $g=\left[\partial\omega_{\mathrm{b}}\left(x\right)/\partial x\right]  x_{\mathrm{ZPF}}$. Here, $x_{\mathrm{ZPF}}=\sqrt{\hbar/(2m_{\mathrm{eff}}\omega_{\mathrm{m}})}$ is the amplitude of the zero-point fluctuation and $m_{\mathrm{eff}}$ is an effective mass of the mechanical resonator. The position coordinate $x$ describes the displacement of the mechanical resonator. The Hamiltonian of the whole system is written as
\begin{equation}
H=\hbar \omega_{\mathrm{a}}a^{\dag }a+\hbar \omega_{\mathrm{b}}b^{\dag}b+\hbar \omega_{\mathrm{m}}c^{\dag}c+\hbar J(a^{\dag }b+b^{\dag }a)+\hbar gb^{\dag}b(c+c^{\dag}).
\label{SPhoton}
\end{equation}

To enhance the optomechanical coupling strength, a control field $S_{in}=\Omega e^{-i\omega_{\mathrm{p}}t}$ with the frequency $\omega_{\mathrm{p}}$ and the amplitude $\Omega$ is applied to coherently drive the passive cavity mode. That is, the driven cavity mode is optomechanically coupled to the mechanical resonator. By rewriting each operator as a sum of its steady-state mean value and a small fluctuation operator, e.g., $a=\alpha _{1}+\delta {\mathrm{a}},\,b=\alpha _{2}+\delta  {\mathrm{b}},\,c=\beta +\delta  {\mathrm{c}}$, and following usual linearized process (see Appendix~\ref{appa}) in the active-cavity coupled optomechanical systems~\cite{phonon,qpt2,stab1,stab2,stab3}, we can derive an effective Hamiltonian of the fluctuation operators
\begin{align}
H_{\mathrm{lin}}  &  =-\hbar\Delta_{\mathrm{a}}a^{\dag}a-\hbar\Delta_{\mathrm{b}}^{^{\prime}}b^{\dag}b+\hbar J(a^{\dag}b+b^{\dag}a)\nonumber\\
& +\hbar\omega_{\mathrm{m}}c^{\dag}c+\hbar\left(G_{\mathrm{lin}}^{\ast}b+G_{\mathrm{lin}}b^{\dag}\right)  \left(c+c^{\dag}\right),\label{Hlin1}
\end{align}
with $\Delta_{\mathrm{a}}=\omega_{\mathrm{p}}-\omega_{\mathrm{a}}$. Here we drop ``$\delta$" for all the fluctuation operators for the sake of simplicity,  e.g., $\delta {\mathrm{a}}\rightarrow a$.  The parameter $\Delta_{\mathrm{b}}^{^{\prime}}=\Delta_{\mathrm{b}}-g\left(\beta+\beta^{\ast}\right)$ is the modified detuning of the cavity mode $b$ with $\Delta_{\mathrm{b}}=\omega_{\mathrm{p}}-\omega_{\mathrm{b}}$.  $G_{\mathrm{lin}}=g\alpha _{2}$ is the coherent-driving-enhanced optomechanical coupling strength. Without loss of generality, hereafter, we assume that $\Delta_{\mathrm{a}}=\Delta_{\mathrm{b}}^{^{\prime}}=\bar{\Delta}$.

The quantum Langevin equations of all fluctuation operators for the linearized Hamiltonian in Eq.~(\ref{Hlin1}) obey
\begin{align}
\dot{a}  &  =\left(  i\bar{\Delta}+\frac{\kappa}{2}\right)  a-iJb-\sqrt{\kappa}a_{\mathrm{in}},\label{Ll1}\\
\dot{b}  &  =\left(  i\bar{\Delta}-\frac{\gamma}{2}\right) b-iJa-iG_{\mathrm{lin}}\left(c+c^{\dag}\right)  -\sqrt{\gamma}b_{\mathrm{in}},\label{Ll2}\\
\dot{c}  &  =\left(  -i\omega_{\mathrm{m}}-\frac{\gamma_{\mathrm{m}}}{2}\right) c-i\left(G_{\mathrm{lin}}^{\ast}b+G_{\mathrm{lin}}b^{\dag}\right)-\sqrt{\gamma_{\mathrm{m}}}c_{\mathrm{in}}. \label{Ll3}
\end{align}
Here  $b_{\mathrm{in}}$ and $c_{\mathrm{in}}$ are the corresponding noise operators with zero mean values and nonzero correlation functions given by
\begin{align}
\left\langle b_{\mathrm{in}}^{\dag }(t)b_{\mathrm{in}}(t^{^{\prime}})\right\rangle  &=0, \label{no1} \\
\left\langle b_{\mathrm{in}}(t)b_{\mathrm{in}}^{\dag }(t^{^{\prime}})\right\rangle  &=\delta (t-t^{^{\prime }}), \label{no2}\\
\left\langle c_{\mathrm{in}}^{\dag }(t)c_{\mathrm{in}}(t^{^{\prime}})\right\rangle  &=n_{\mathrm{th}}\delta (t-t^{^{\prime }}),\label{no3}\\
\left\langle c_{\mathrm{in}}(t)c_{\mathrm{in}}^{\dag }(t^{^{\prime}})\right\rangle  &=(n_{\mathrm{th}}+1)\delta (t-t^{^{\prime }}),
\label{no4}
\end{align}
with the thermal phonon number $n_{\mathrm{th}}$ of the mechanical resonator
\begin{equation}
n_{\mathrm{th}}=(\mathrm{e}^{\hbar \omega _{\mathrm{m}}/k_{\mathrm{B}}T}-1)^{-1},\label{nth1}
\end{equation}
where $T$ is the environmental temperature and $k_{\mathrm{B}}$ is the Boltzmann constant. Note that the realizations of the gain in optical frequency domain are always associated  with laser amplification with the population inversion of the gain system, e.g., the two-level systems~\cite{laser1,laser2}. For the active cavity mode, the intrinsic quantum noise is described by the noise operators $a_{\mathrm{in}}$ and $a^{\dagger}_{\mathrm{in}}$, which satisfy~\cite{qpt1,qpt2,qpt3,qpt4,qpt5}
\begin{align}
\left\langle a_{\mathrm{in}}(t)a_{\mathrm{in}}^{\dag }(t^{^{\prime}})\right\rangle  &=0, \label{ano1}\\
\left\langle a_{\mathrm{in}}^{\dag }(t)a_{\mathrm{in}}(t^{^{\prime}})\right\rangle  &=\delta (t-t^{^{\prime}}).\label{ano2}
\end{align}
In the above, we have assumed that the optical frequencies is very high such that the temperature effect on the cavity modes is negligibly small. That is, the environments of two cavities are assumed to be in vacuum state, thus the correlation functions of the cavity noise operators are independent of the environmental temperature $T$.

\section{Giant cooling rate}
Let us now study the cooling rate of the mechanical resonator by using the quantum noise spectrum of the optical force. We know that the force acting on the mechanical resonator can be obtained by $-\partial H/\partial x$ with $x=x_{\rm {ZPF}}(c^\dagger+c)$. From Eq.~(\ref{Hlin1}), we can derive the optical force operator
\begin{equation}
F(t)=-\frac{\hbar}{x_{\mathrm{ZPF}}}\left[G_{\mathrm{lin}}^{\ast}b(t)+G_{\mathrm{lin}}b(t)^{\dag}\right]. \label{OpFo}
\end{equation}
The quantum noise spectrum of the optical force is defined by the Fourier transform of the autocorrelation function, i.e.,
\begin{equation}
S_{\mathrm{FF}}(\omega)=\int dt\,e^{i\omega t}\langle F(t)F(0)\rangle.
\end{equation}

To obtain the expression for $S_{\mathrm{FF}}(\omega)$, we can transform Eqs.~(\ref{Ll1})-(\ref{Ll3}) to the frequency domain, and have
\begin{align}
\frac{a\left( \omega \right) }{\chi _{\mathrm{a}}\left( \omega \right)}
&=-iJb\left( \omega \right)-\sqrt{\kappa }a_{\mathrm{in}}\left( \omega\right) ,  \label{Lw1} \\
\frac{b\left( \omega \right) }{\chi _{\mathrm{b}}\left( \omega \right)}&=-iJa\left( \omega \right)-iG_{\mathrm{lin}}\left[ c\left( \omega \right)+c^{\dag }\left( \omega \right) \right] -\sqrt{\gamma }b_{\mathrm{in}}\left(\omega \right) ,  \label{Lw2} \\
\frac{c\left( \omega \right) }{\chi _{\mathrm{c}}\left( \omega \right)}& =-i\left[ G_{\mathrm{lin}}^{\ast }b\left( \omega \right)+G_{\mathrm{lin}}b^{\dag }\left( \omega \right) \right] -\sqrt{\gamma _{\mathrm{m}}}c_{\mathrm{in}}\left( \omega \right), \label{Lw3}
\end{align}
where the response functions $\chi_{\mathrm{a}}\left(\omega \right)$, $\chi _{\mathrm{b}}\left(\omega \right)$, and $\chi _{\mathrm{c}}\left(\omega \right)$ of the active cavity mode, passive cavity mode, and mechanical mode are given by
\begin{align}
\chi_{\mathrm{a}}\left( \omega \right) & =\frac{1}{-i\left(\omega+\bar{\Delta}\right)-\kappa /2 },  \label{COOL6} \\
\chi_{\mathrm{b}}\left( \omega \right) & =\frac{1}{-i\left(\omega+\bar{\Delta}\right)+\gamma /2 },  \label{COOL7}\\
\chi_{\mathrm{c}}\left( \omega \right) & =\frac{1}{-i\left(\omega-\omega_{\mathrm{m}}\right)+\gamma_{\mathrm{m}}/2}.  \label{COOL8}
\end{align}
By using $F(\omega)=-\hbar\left[G_{\mathrm{lin}}^{\ast }b\left(\omega \right)+G_{\mathrm{lin}}b^{\dag}\left(\omega \right) \right]/x_{\mathrm{ZPF}}$ in the frequency domain, the quantum noise spectrum for our system is  given (see Appendix~\ref{appb} for detailed derivations)
\begin{equation}
S_{FF}\left(\omega \right) =\frac{\hbar^{2}\left\vert G_{\mathrm{lin}}\right\vert^{2}}{x_{\mathrm{ZPF}}^{2}}\left[ \gamma \left\vert \chi \left( \omega\right) \right\vert ^{2}+J^{2}\kappa \left\vert \chi_{\mathrm{a}}\left(-\omega \right) \right\vert ^{2}\left\vert \chi \left( -\omega \right)\right\vert^{2}\right], \label{SFFW}
\end{equation}
where
\begin{equation}
\chi \left( \omega \right) =\frac{\chi_{\mathrm{b}}\left(\omega \right)}{1+\left\vert G_{\mathrm{lin}}\right\vert^{2}\chi _{\mathrm{c}}\left(\omega \right) \chi_{\mathrm{b}}\left(\omega \right)+J^{2}\chi_{\mathrm{a}}\left(\omega \right) \chi _{\mathrm{b}}\left(\omega \right)}, \label{chit}
\end{equation}
represents the total response function of the coupled active and passive cavities.
\begin{figure}[ptb]
\includegraphics[bb=0 215 560 610,  width=8cm, clip]{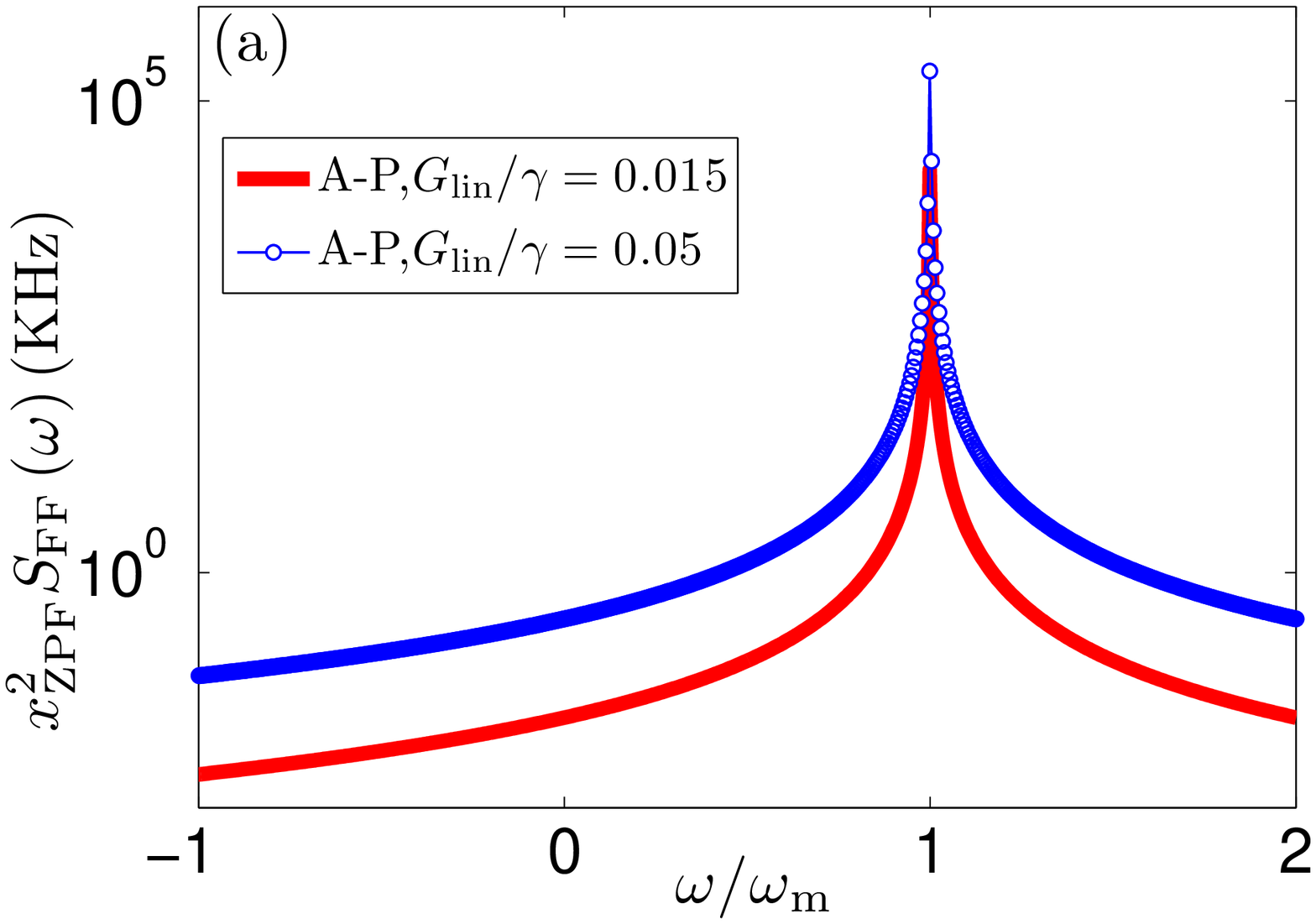}
\includegraphics[bb=0 215 560 610,  width=8cm, clip]{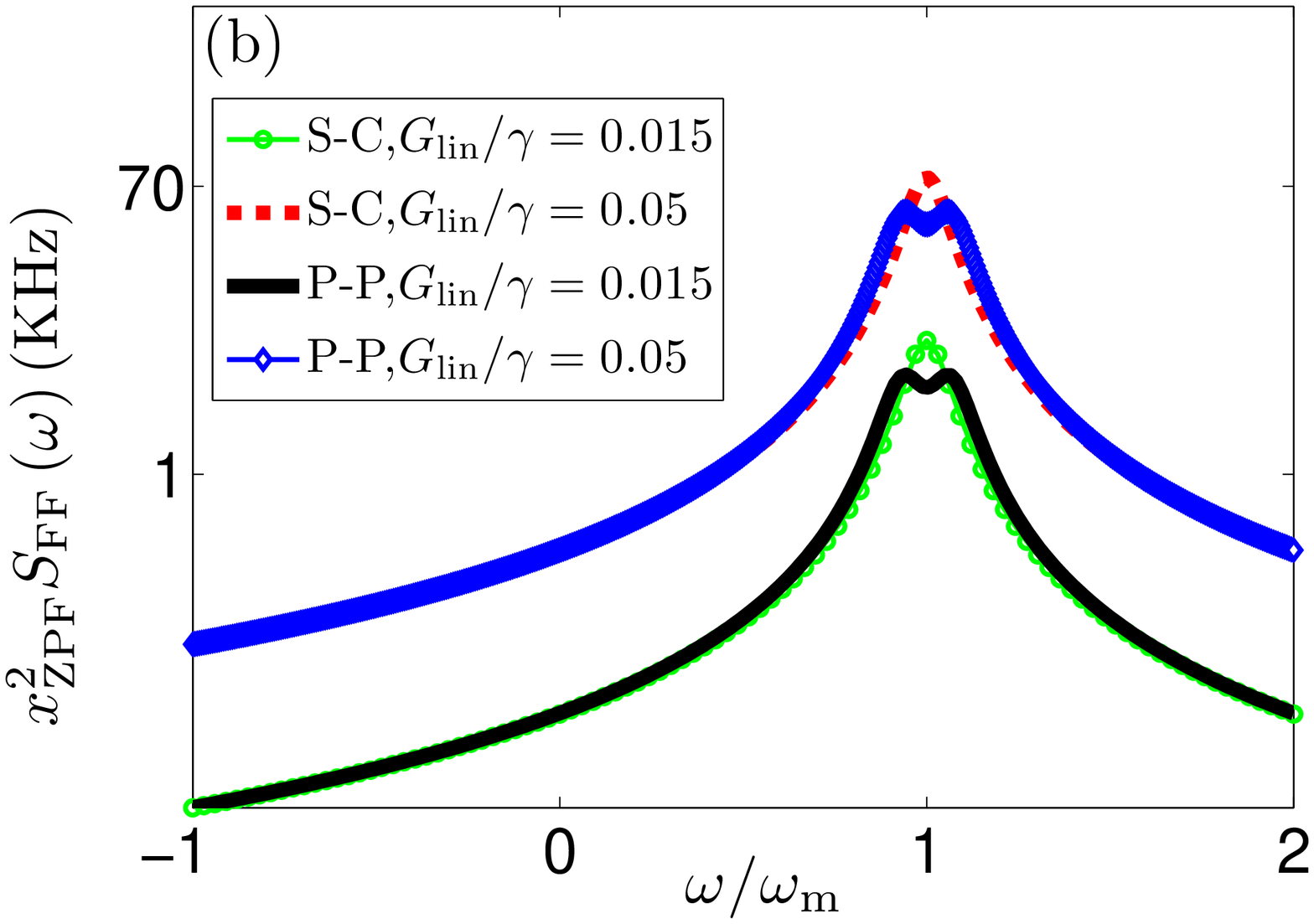}
\caption{(Color online) Optical force spectrum $S_{\mathrm{FF}}\left(\omega\right)  $ versus $\omega$ for cases: (a) active-cavity (A-P) assisted optomechanical system; (b) standard optomechanical system (S-C), and passive-cavity (P-P) assisted optomechanical systems. Two different coupling strengths $G_{\mathrm{lin}}/\gamma=0.015$ and $0.05$, are considered. Here, we set $\kappa=\gamma$ for the A-P case and $\kappa=-\gamma$ for the P-P case, respectively. Other parameters are: $\omega_{\mathrm{m}}/2\pi=20$ $\mathrm{MHz}$, $\gamma_{\mathrm{m}}/\omega_{\mathrm{m}}=10^{-5}$, $\gamma/\omega_{\mathrm{m}}=1/\left(  5\pi\right)  $, $J/\gamma=0.49$, and $\bar{\Delta}\equiv\omega_{\mathrm{p}}-\omega_{\mathrm{a}}=-\omega_{\mathrm{m}}$.}\label{fig2}
\end{figure}

The photon number fluctuations give rise to a noisy force on the mechanical resonator, which cause transitions between phonon states of the mechanical resonator. Using the Fermi golden rule, the cooling and heating rates can be calculated and given by $A_{-}\equiv S_{\mathrm{FF}}\left(\omega_{\mathrm{m}}\right) x_{\mathrm{ZPF}}^{2}/\hbar^{2}$, and $A_{+}\equiv S_{\mathrm{FF}}\left(-\omega_{\mathrm{m}}\right)x_{\mathrm{ZPF}}^{2}/\hbar^{2}$, respectively (the detailed derivations can be found in Appendix \ref{appB}). To cool down the mechanical resonator, the cooling rate $A_{-}$ should be larger than the heating rate $A_{+}$. In Fig.~\ref{fig2}(a), the optical force spectrum $S_{\mathrm{FF}}(\omega)$ is plotted as a function of $\omega$ for the active-cavity assisted optomechanical system (represented by A-P). For comparisons, the optical force spectra $S_{\mathrm{FF}}(\omega)$ are also shown in Fig.~\ref{fig2}(b) for a standard optomechanical system (represented by S-C) or a system (represented by P-P) that the cavity of the standard optomechanical system is coupled to an additional loss cavity (we call it as passive-cavity assisted optomechanical system). We emphasize that the driven cavity mode is optomechanically coupled to the mechanical resonator for all S-C, P-P and A-P cases.
\begin{figure}[ptb]
\includegraphics[bb=140 300 430 535,  width=9cm, clip]{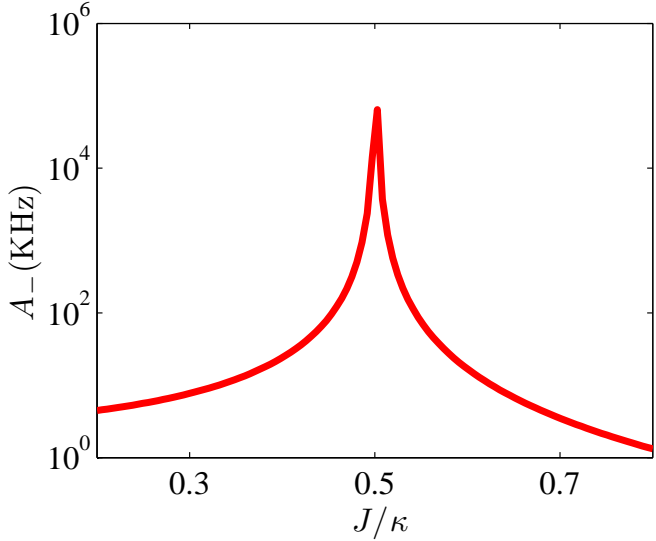}\caption{(Color online) The cooling rate $A_{-}$ versus coupling strength $J$ between the two coupled optical cavities are calculated. Here, the gain and decay rates are balanced, i.e., $\kappa=\gamma$, and the other parameters are the same as those in Fig.~\ref{fig2}.}
\label{fig3}
\end{figure}

For all three cases with experimentally accessible parameters~\cite{parameter}, Fig.~\ref{fig2} shows that the red detuning of the driving field to the cavity field (i.e., $\bar{\Delta}\equiv\omega_{\mathrm{p}}-\omega_{\mathrm{a}}<0$) leads to $A_{-}>$ $A_{+}$, corresponding to the cooling. We also find that a larger optomechanical coupling strength $G_{\mathrm{lin}}$ corresponds to a larger cooling rate. The maximum value of the optical force spectrum $S_{\mathrm{FF}}\left(\omega\right)$ locates at $\omega=\omega_{\mathrm{m}}$, corresponding to the maximum cooling rate for the A-P and S-C cases. However, as shown in Fig.~\ref{fig2}(b), for the P-P case,  the optical force spectrum $S_{\mathrm{FF}}\left(\omega\right)$ has two peaks which originate from the mode coupling between two passive optical modes. Such spectrum splitting makes the optimal cooling frequency shift, and also reduces the maximum cooling rate compared to the A-P and S-C cases. For the A-P case, the maximum value of the optical force spectrum is enhanced about four orders of magnitude at $\omega=\omega_{\mathrm{m}}$ compared to those for both the S-C and the P-P cases. Thus a giant enhancement of the cooling rate $A_{-}$ can be obtained in our proposed active-cavity assisted optomechanical system. This is very different from the P-P case where the mode coupling reduces the cooling rate.

We find that the optimized cooling rate can be obtained when the optical gain and loss are balanced, i.e., $\kappa=\gamma$. As shown in Fig.~\ref{fig3}, the optimized cooling rate appears at $J=\kappa/2=\gamma/2$, corresponding to the maximum value of $A_{-}$. This parameter condition for the optimal cooling rate can also be obtained from its analytical expression. Under the weak optomechanical coupling condition and the condition that the gain and loss are balanced, i.e., $G_{\mathrm{lin}}/\gamma\ll1$ and $\kappa=\gamma$, the cooling rate in Eq.~(\ref{SFFW}) can be further simplified to
\begin{equation}
A_{-}{=\left\vert G_{\mathrm{lin}}\right\vert^{2}}\frac{\gamma ^{3}}{\left[J^{2}-\left( \gamma /2\right) \right] ^{2}}+o(\gamma ). \label{seq}
\end{equation}
where $o(\gamma)$ is a small quantity  which can be neglected in the following discussions. Equation~(\ref{seq}) clearly shows that the maximum value of $A_{-}$ (optimal cooling rate) locates at $J=\gamma/2=\kappa/2$. It is notable that when the optical gain and loss are set to be balanced and two cavities are degenerated, the coupled optical cavities are $\mathcal{PT}$-symmetric~\cite{PT2,PT2s,phonon,qpt2,stab1,stab2,stab3}. Here, the optimal point for maximum cooling rate corresponds to the phase transition point of the $\mathcal{PT}$-symmetry. In the following section, we will analyze the relation between the phase transition and the cooling rate of the mechanical resonator.

\begin{figure*}[ptb]
\includegraphics[bb=0 200 560 620,  width=8.2cm, clip]{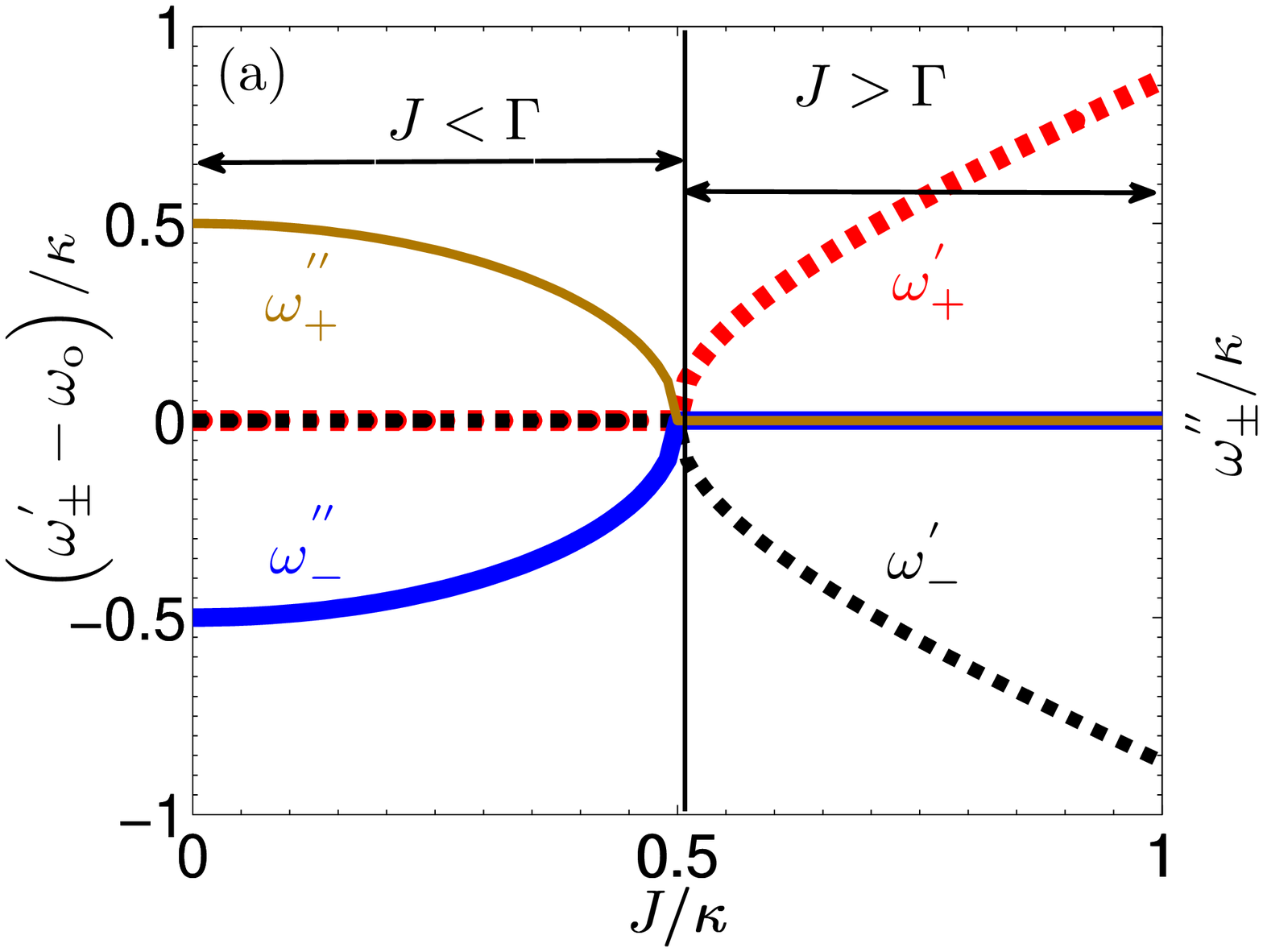}\includegraphics[bb=0 200 560 620,  width=8.3cm, clip]{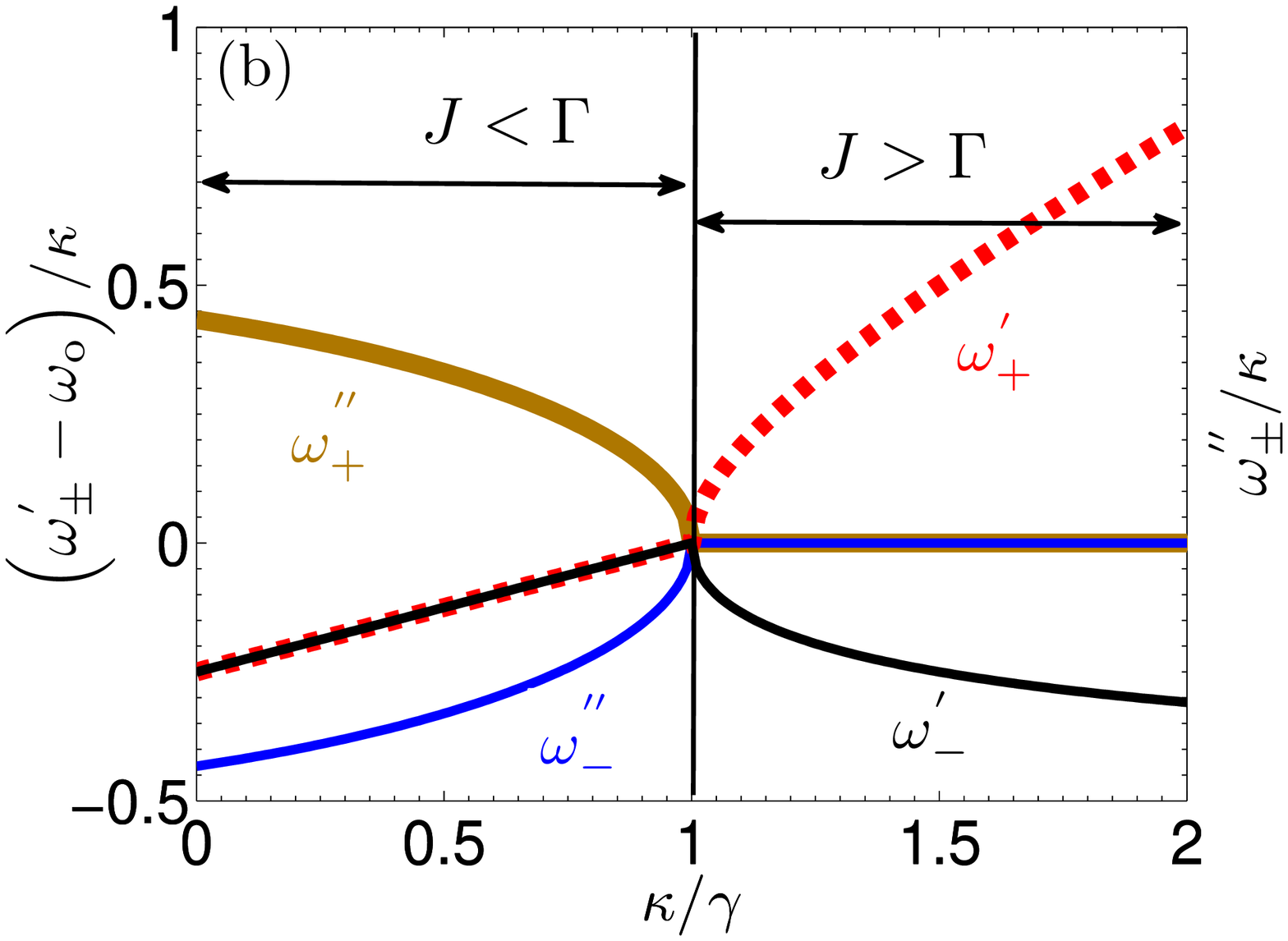}
\includegraphics[bb=0 150 560 350,  width=16cm, clip]{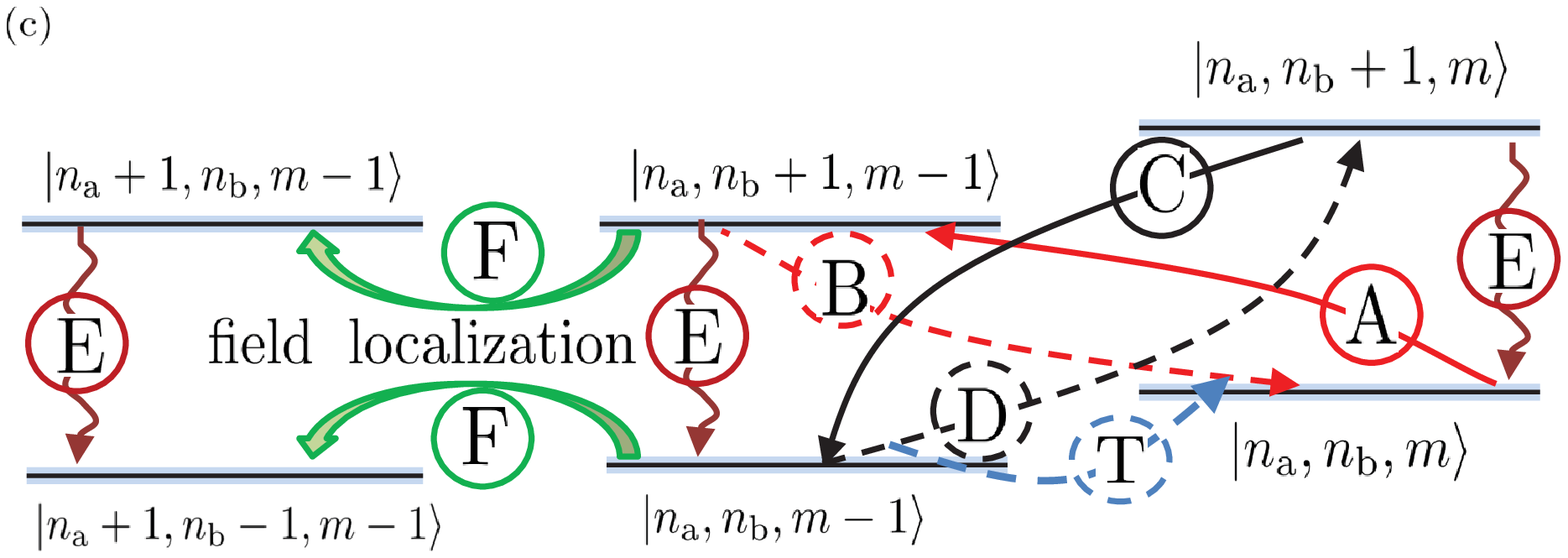}
\caption{(Color online) Phase transitions induced by increasing coupling strength $J$ with balanced gain and decay rates of the optical modes (i.e., $\kappa/\gamma=1$), and by increasing gain $\kappa$ with $J/\gamma=0.5$, are shown in (a) and (b), respectively. Here, the real and imaginary parts of complex numbers $\omega_{\pm}$ are denoted by $\omega_{\pm}^{^{\prime}}$ (denoted by the red- and black-dashed line) and $\omega_{\pm}^{^{\prime\prime}}$ (denoted by the brown- and blue-solid line), respectively; (c) Energy levels diagram for the cooling mechanism. $\left\vert n_{\mathrm{a}},n_{\mathrm{b}},m\right\rangle $ denotes the number state with $n_{\mathrm{a} }$ ($n_{\mathrm{b}}$) being the photons number of the active (passive) optical mode $a$ ($b$), and $m$ being the phonon number of the mechanical mode. Three kinds of heating processes are denoted by\ process B (swap heating), D (quantum backaction heating), and T (thermal heating), respectively. The cooling processes are represented by A (swap cooling), C (counter-wave cooling), E (dissipation cooling by optical mode $b$), and F (field localization cooling by optical mode $a$), respectively.}
\label{fig4}
\end{figure*}

\section{phase transition and optomechanical cooling}

If we only consider two coupled cavities, we can write out an effective Hamiltonian
\begin{equation}
H_{\mathrm{c}}=\hbar\omega_{\mathrm{a}}a^{\dag}a+\hbar\omega_{\mathrm{b}}b^{\dag}b+i(\kappa/2)a^{\dag}a-i(\gamma/2)b^{\dag}b
+\hbar J(a^{\dag}b+b^{\dag}a). \label{effech}
\end{equation}
by including the gain (loss) rate of active (passive) mode $a$ ($b$). When two cavities are degenerated, i.e., $\omega_{\mathrm{1}}=\omega_{\mathrm{2}}=\omega_{\mathrm{0}}$, we can further rewrite the Hamiltonian $H_{\mathrm{c}}$ in Eq.~(\ref{effech}) as
\begin{equation}
H_{\mathrm{c}}=\hbar\omega_{+}A_{1}^{\dagger}A_{1}+\hbar\omega_{-}A_{2}^{\dagger}A_{2},
\end{equation}
where two supermode operators are defined as $A_{1}=\left(b+a\right)/\sqrt{2}$, and $A_{2}=\left(b-a\right)/\sqrt{2}$ with the frequencies
\begin{equation}
\omega_{\pm}=\omega_{\mathrm{0}}-i\frac{\chi}{2}\pm\sqrt{J^{2}-\Gamma^{2}},\label{supermode}
\end{equation}
where $\chi=(\gamma-\kappa)/2$ and $\Gamma=(\kappa+\gamma)/4$. The supermode $A_{1}$ ($A_{1}$) corresponds to the frequency $\omega_{+}$ ($\omega_{-}$). If the Hamiltonian of two coupled cavities is unchanged under both the time- and parity- reversal transformation, we call it as a $\mathcal{PT}$-symmetric system. Here, to realize the $\mathcal{PT}$-symmetry, we have assumed that two prerequisites are satisfied: (i) the gain rate of the active mode $a$ and the decay rate of the passive mode $b$ are balanced, i.e., $\kappa=\gamma$; and (ii) two cavity modes are degenerated, i.e., $\omega_{\mathrm{1}}=\omega_{\mathrm{2}}=\omega_{\mathrm{0}}$.
\begin{figure*}[ptb]
\includegraphics[bb=0 190 570 670,  width=6cm, clip]{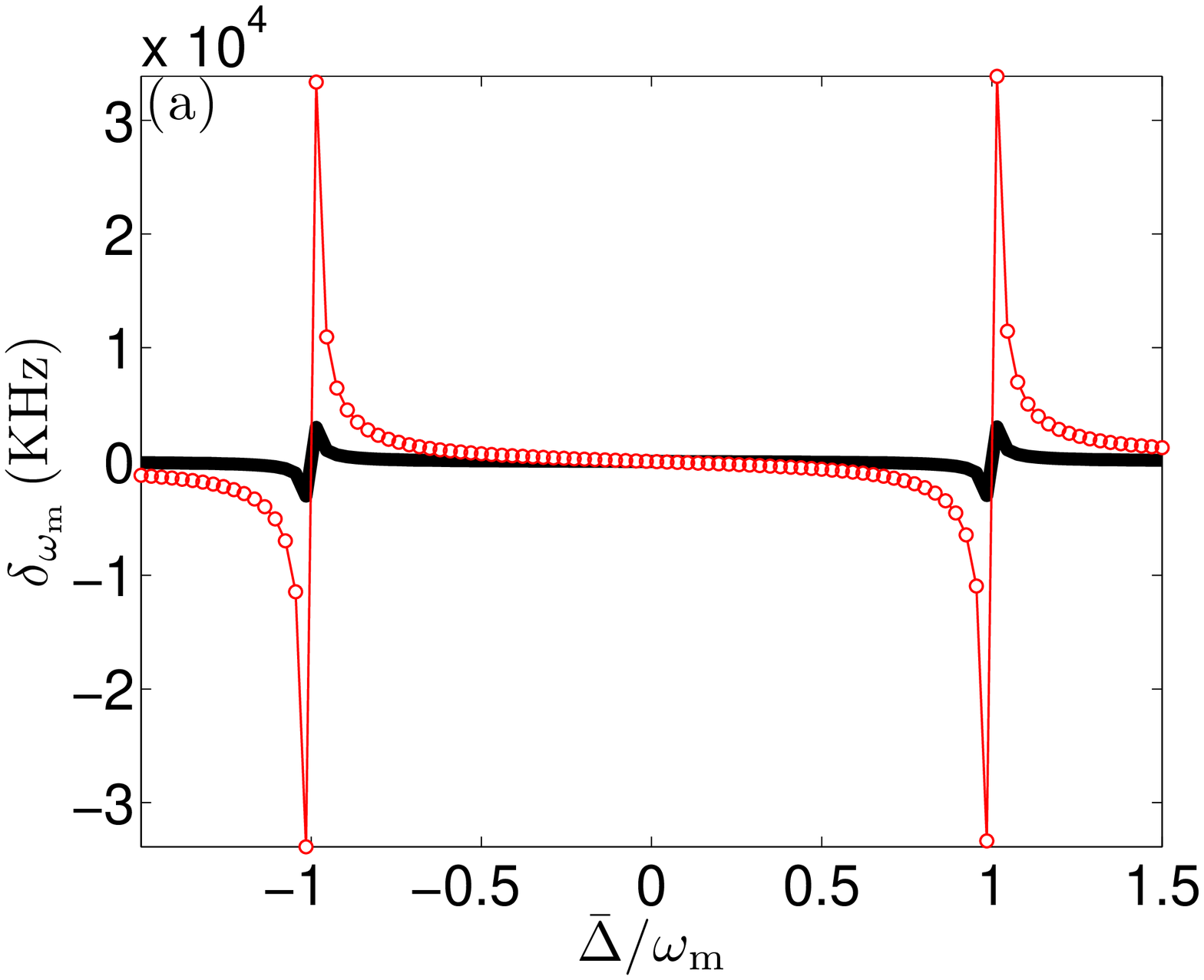}\includegraphics[bb=0 190 570 670, width=6cm, clip]{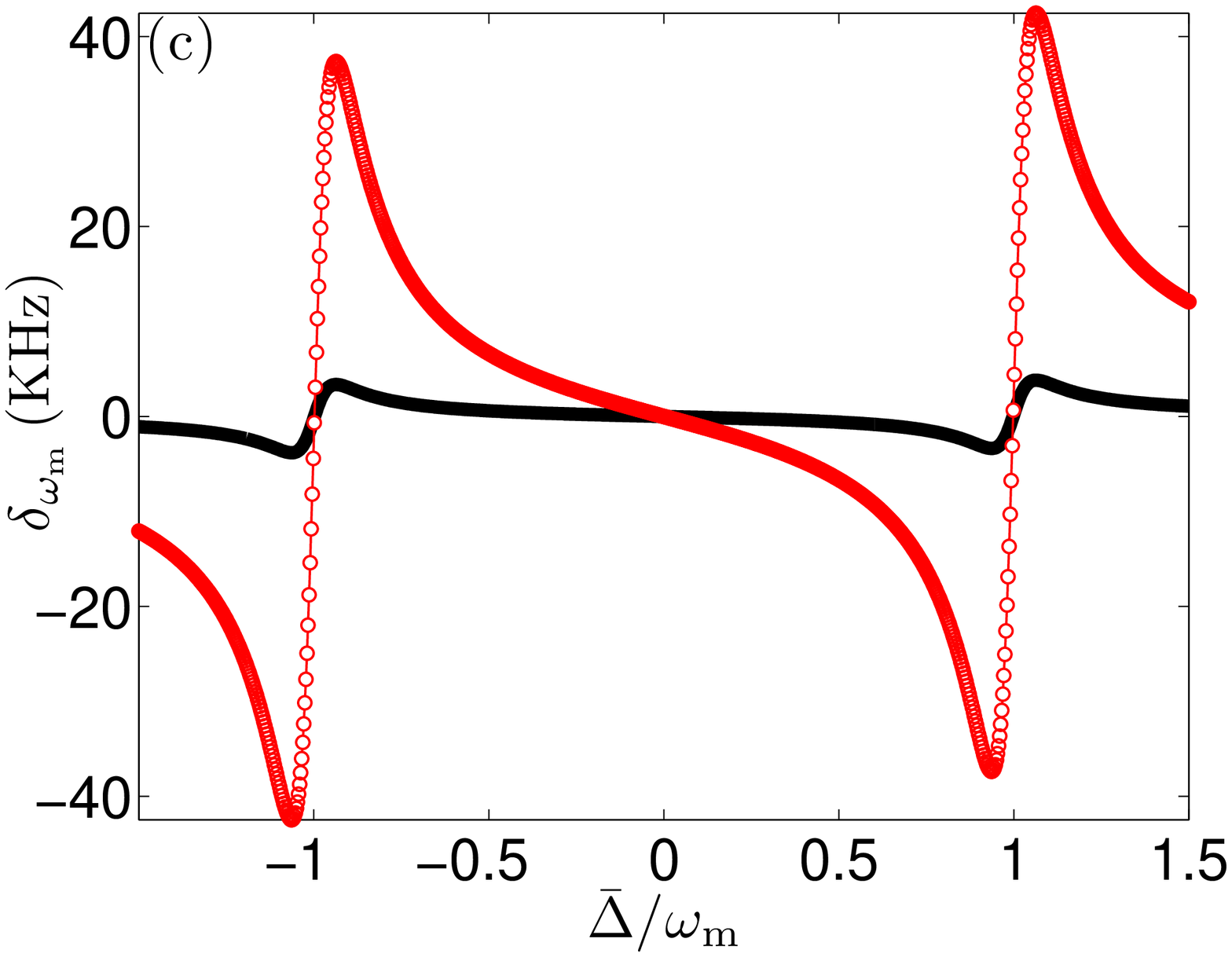}\includegraphics[bb=0 190 570 670,  width=6cm, clip]{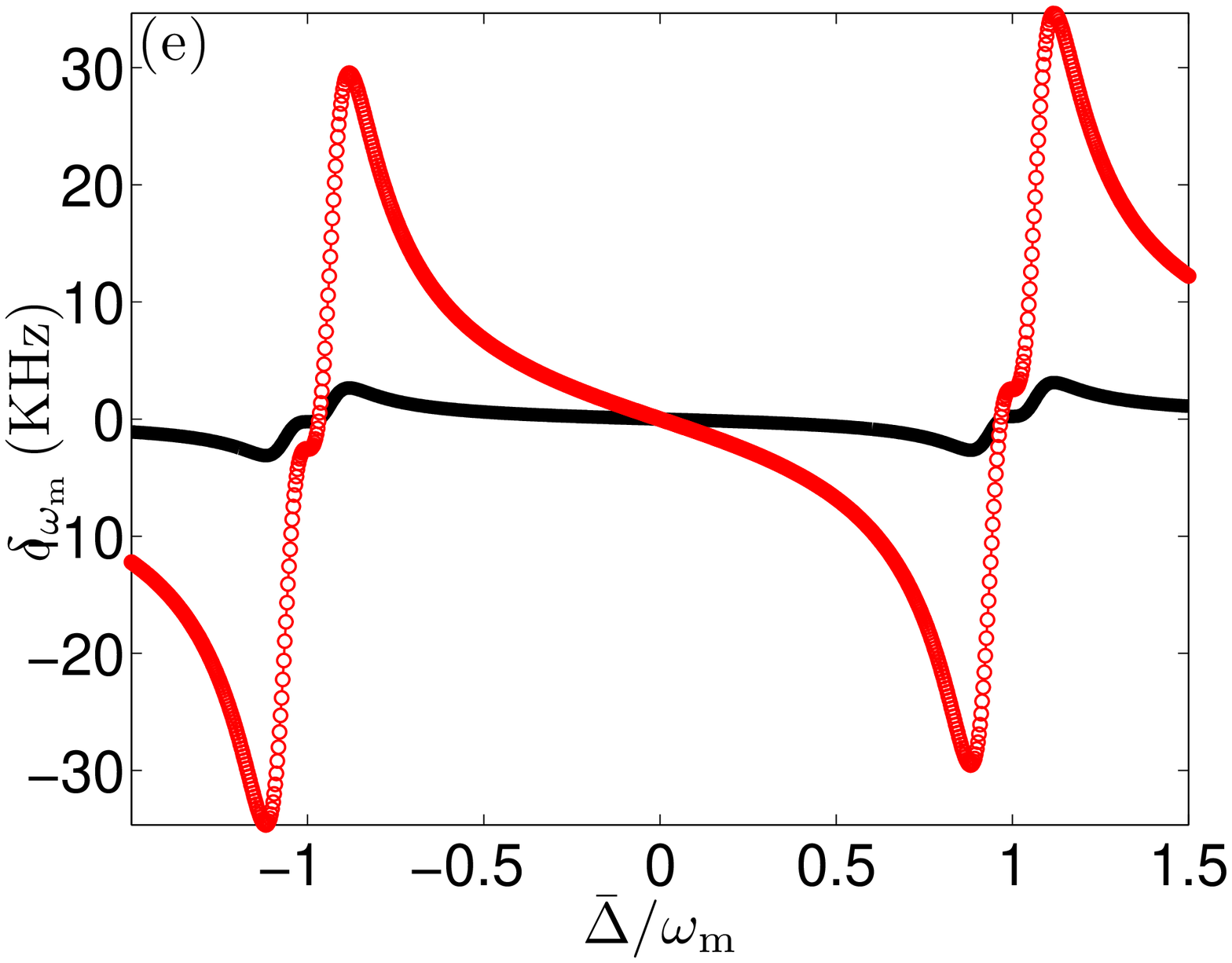}
\includegraphics[bb=0 190 570 670,  width=6cm, clip]{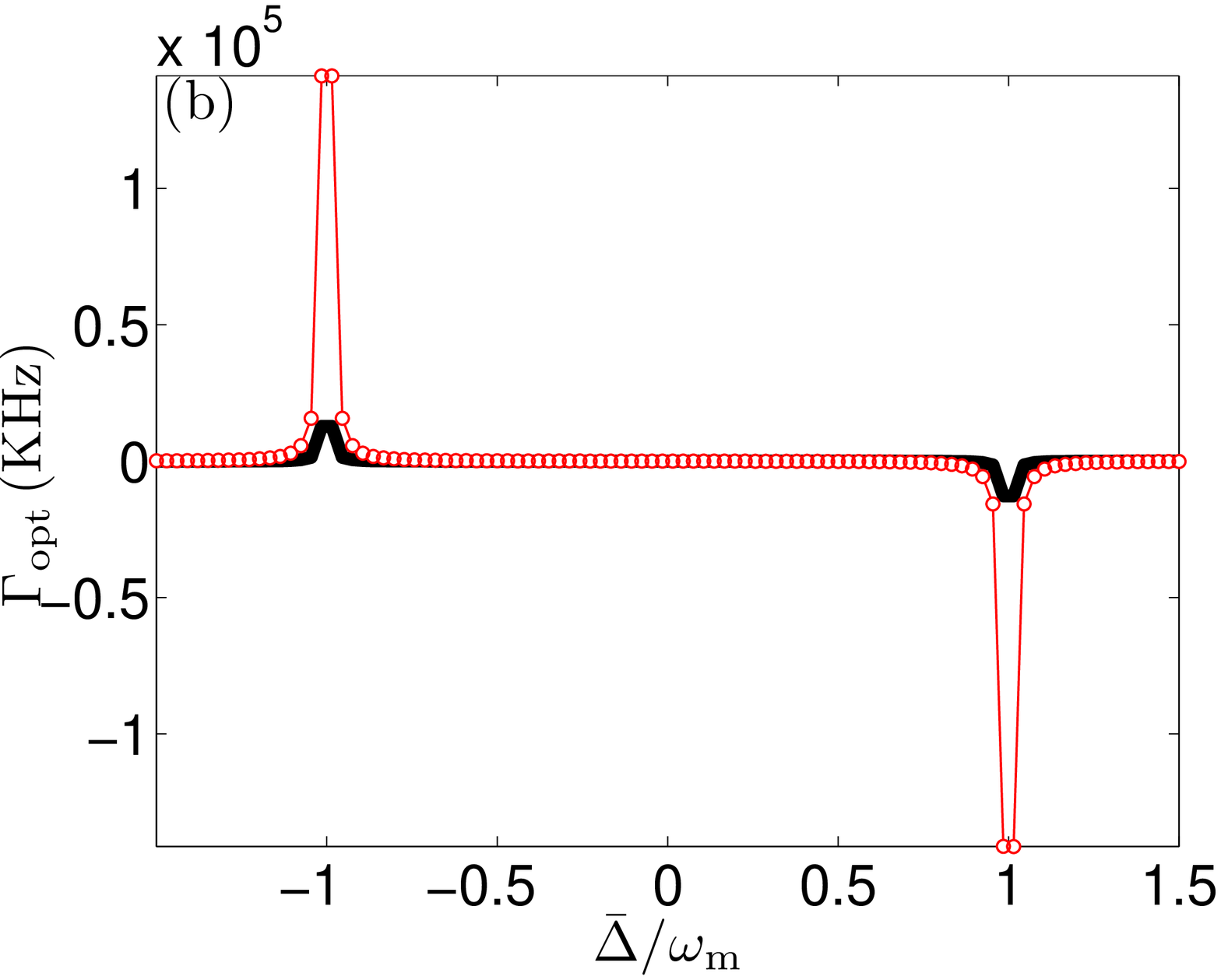}\includegraphics[bb=0 195 570 630,  width=6cm, clip]{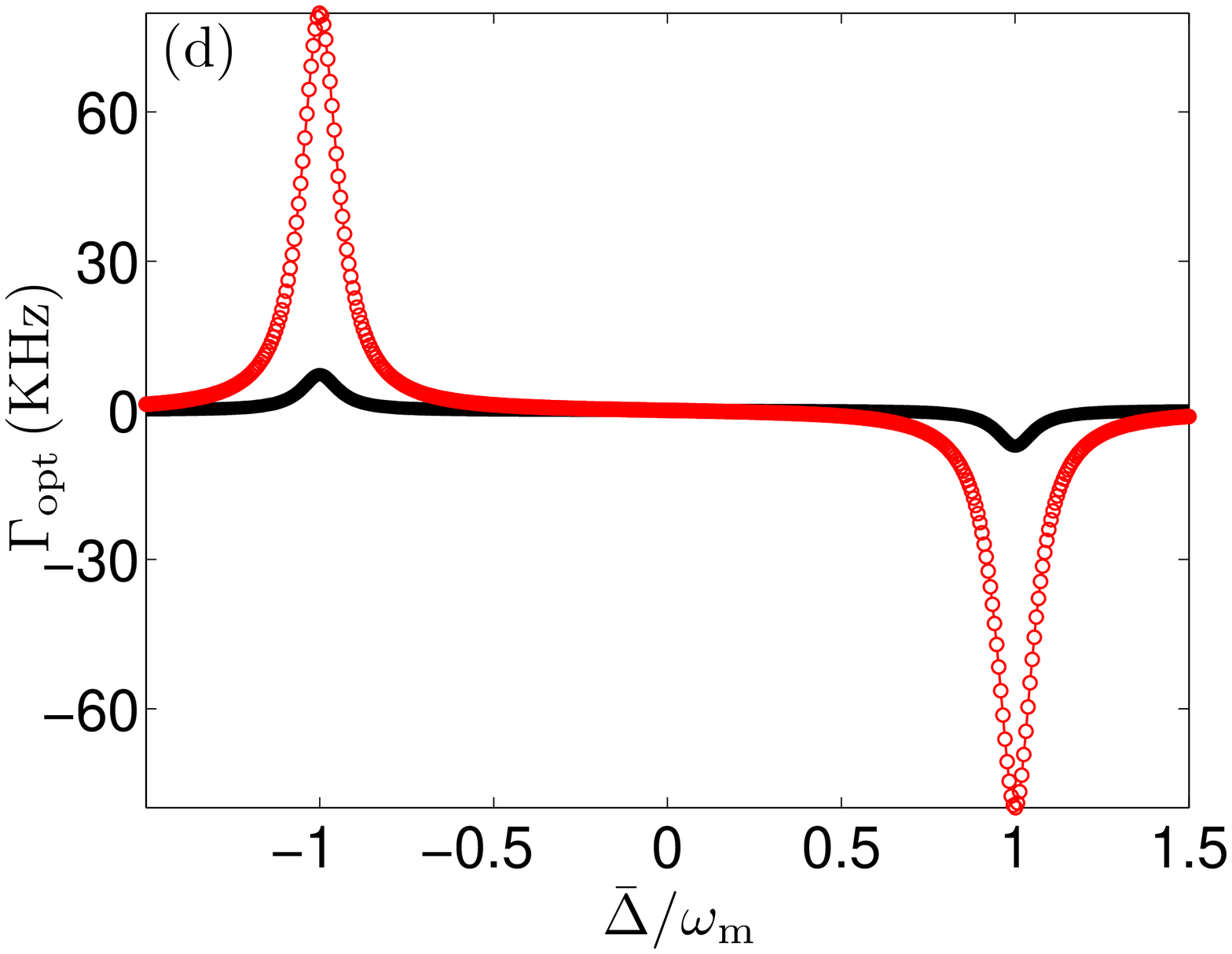}\includegraphics[bb=0 190 570 670,  width=6cm, clip]{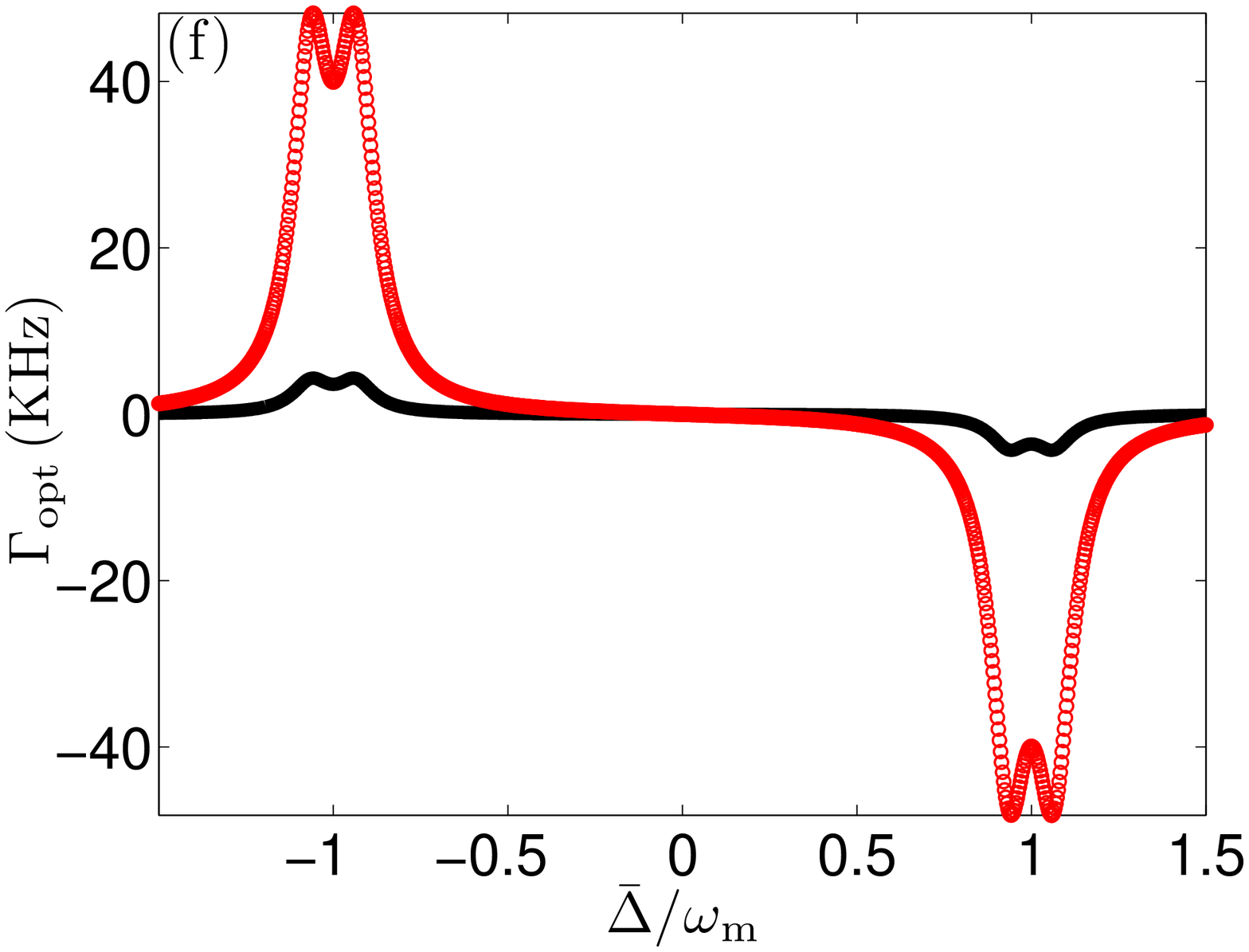}
\caption{(Color online) Frequency shift $\delta_{\mathrm{\omega}_{\mathrm{m}}}$, and optical damping $\Gamma_{\mathrm{opt}}$ versus $\bar{\Delta}$ with two coupling strengths $G_{\mathrm{lin}}/\gamma=0.015$ (black curve) and $0.05$ (read dotted curve), for A-P case (shown in (a) and (b)), S-C case (shown in (c) and (d)), and P-P case (shown in (e) and (f)). Other parameters are the same as those in Fig.~\ref{fig2}, except for $\omega=\omega_{\mathrm{m}}$.}
\label{fig5}
\end{figure*}

From Eq.~(\ref{supermode}) and also as shown in Fig.~\ref{fig4}(a), there are mainly two parameter regimes to characterize $\omega_{\pm}$. When the mode-coupling strength $J$ is larger than the effective loss $\Gamma$ of the supermodes, e.g., $J>\Gamma$ ($J>\gamma/2$ or $J>\kappa/2$ for $\kappa=\gamma$), two supermodes have different frequencies but with the same linewidth $\chi$, which approaches to zero when the gain and loss are comparable, i.e., $\kappa\approx\gamma$. However, when $J<\Gamma$ ($J<\gamma/2$ or $J<\kappa/2$ for $\kappa=\gamma$), two supermodes become degenerate but with different linewidths. Here, $J=\Gamma$ ($J=\gamma/2$ or $J=\kappa/2$ for $\kappa=\gamma$) corresponds to the phase transition point of the $\mathcal{PT}$-symmetric systems. It is obvious that the dynamics of two supermodes can be changed from the same decay rate for both modes to that one has loss and other one has gain by either decreasing the mode-coupling strength $J$ as shown in Fig.~\ref{fig4}(a), or increasing the gain rate $\kappa$ in the active cavity as shown in Fig.~\ref{fig4}(b). Here, we assume that the decay rate of the passive cavity is given when the sample is fabricated.

Around the phase transition point, it was found that one supermode experiences gain (active supermode) and other one has loss (passive supermode), thus the system is not in equilibrium because the energy of the active supermode exponentially grows, but the energy of the passive supermode exponentially decays. As a result, the optical field is strongly localized in the active cavity. The field localization has been experimentally observed in the so-called $\mathcal{PT}$-symmetric systems~\cite{PT2,PT2s}, and is further explored to realize the low-power photon diode~\cite{PT2}, single-mode lasers~\cite{PT3,PT4}, a giant phonon nonlinearity~\cite{phonondiode,jingp}, a ultralow-power chaos~\cite{chaos}, and a unconventional EIT~\cite{liuyulong}.

From Eq.~(\ref{seq}) and as also shown in Fig.~\ref{fig3}, we find that the optimized cooling rate locates at the phase transition point of the $\mathcal{PT}$-symmetric optical cavities. Thus, the energy localization effect at the phase transition point can be used to enhance the cooling rate of the mechanical resonator. The physical mechanism is schematically shown in Fig.~\ref{fig4}(c). The field localization can greatly enhance the absorption rate of the anti-Stokes photons (process F) of the optical mode $b$, compared to the standard optomechanical case for which there is only cavity dissipation (process E). Simultaneously, the rate of  heating exchange (process B) is also greatly reduced due to the field localization and the decrease of the anti-Stokes photons. These finally enable a dynamical enhancement of the cooling rate of phonons.

We emphasize that the correspondence between the optimal point and the phase transition point is obtained under the condition that our system works in the weak optomechanical coupling regime, i.e.,  $G_{\mathrm{lin}}/\gamma\ll1$, where the mechanical effect on the phase transition point for our three-mode system, represented by the Hamiltonian in Eq.~(\ref{Hlin1}), can be safely neglected. However, when the optomechanical coupling strength is strong enough, e.g., in the strong coupling regime where $G_{\mathrm{lin}}/\gamma>1$, the mechanical effects on the phase transition points cannot be neglected. We should consider the whole three-mode case, consisting of an active cavity coupled to a passive cavity supporting a mechanical mode. The optimal point for the maximum cooling rate is also changed to
\begin{equation}
J=\sqrt{\left(\frac{\gamma}{2}\right)^{2}+4\left\vert G_{\mathrm{lin}}\right\vert ^{2}\frac{\gamma}{\gamma _{\mathrm{m}}}}.
\end{equation}
The emergence and exotic effects of the third-order exceptional points and the relation between the third-order exceptional points and the optimal point of the cooling rate is further explored in a recent work~\cite{horderep}.

\section{Active-cavity assisted optical spring effect}

Let us now explore the frequency shift $\delta_{\mathrm{\omega}_{\mathrm{m}}}$ and the net optical damping  $\Gamma_{\mathrm{opt}}$. From solution of $c(\omega)$ given in Eq.~(\ref{COOL1}),  we know that the real part of $\sum(\omega_{\mathsf{m}})$ corresponds to the frequency shift $\delta_{\mathrm{\omega}_{\mathrm{m}}}$ of the mechanical resonator induced by the optomechanical coupling, i.e.,
\begin{equation}
\delta_{\mathrm{\omega}_{\mathrm{m}}}=\operatorname{Re}[\sum(\omega_{\mathsf{m}})].
\end{equation}
The imaginary part of  $\sum(\omega_{\mathsf{m}})$ corresponds to the net optical damping rate of the mechanical mode induced by the optomechanical coupling, i.e.,
\begin{equation}
\Gamma_{\mathrm{opt}}=-2\operatorname{Im}[\sum(\omega_{\mathsf{m}})]=A_{-}-A_{+}.
\end{equation}
Detailed expressions and  $\sum(\omega_{\mathsf{m}})$ can be found in Appendix~\ref{appc}.

In Fig.~\ref{fig5}, both the frequency shift $\delta_{\omega_{\mathrm{m}}}$ and the optical damping $\Gamma_{\mathrm{opt}}$ are plotted as a function of the detuning $\bar{\Delta}=\omega_{p}-\omega_{a}$ between the driving field and the cavity mode for all three cases (A-P, S-C, and P-P)  with two coupling strengths, i.e., $G_{\mathrm{lin}}/\gamma=0.015$ and $0.05$, respectively.
For the A-P case, as shown in Figs.~\ref{fig5}(a) and (b), the maximum frequency shift occurs around the point $\bar{\Delta}=\pm\omega_{\mathrm{m}}$; the maximum optical damping (gain) locates at the point $\bar{\Delta}=-\omega_{\mathrm{m}}$ ($\omega_{\mathrm{m}}$), at which the frequency shift equals to zero. These are similar to those observed for the S-C case, as shown in Figs.~\ref{fig5}(c) and (d). The optical spring effect of the mechanical resonator (including the frequency shift $\delta_{\omega_{\mathrm{m}}}$ and the optical damping $\Gamma_{\mathrm{opt}}$) for the S-C case have been observed in experiments~\cite{bc1,bc2,bc3,bc4}. Compared to the S-C case, Figs.~\ref{fig5}(a) and (b) show the significant enhancements of the frequency shift $\delta_{\omega_{\mathrm{m}}}$ and the optical damping $\Gamma_{\mathrm{opt}}$ for the A-P case.

At the phase transition point, the net optical damping for mechanical resonator is given as
\begin{equation}
\Gamma _{\mathrm{opt}}=\kappa \left\vert G_{\mathrm{lin}}\right\vert ^{2}\left[ \frac{1}{\left(\omega_{\mathrm{m}}+\bar{\Delta} \right) ^{2}}-\frac{1}{\left( \bar{\Delta} -\omega _{\mathrm{m}}\right) ^{2}}\right]. \label{neto}
\end{equation}
It clearly shows that the maximum value corresponding to the largest optical damping locates at point $\bar{\Delta}=-\omega_{\mathrm{m}}$, and the minimum value corresponding to the largest optical gain for the mechanical mode locates at point $\bar{\Delta}=\omega_{\mathrm{m}}$. These findings have also been shown in Fig.~\ref{fig5}(b).

For the P-P case, as shown in Fig.~\ref{fig5}(e), although the maximum frequency shift $\delta_{\omega_{\mathrm{m}}}$ also occurs around the point $\bar{\Delta}=\pm\omega_{\mathrm{m}}$, but as shown in Fig.~\ref{fig5}(f), the optomechanically induced damping rate $\Gamma_{\mathrm{opt}}$ splits into two peaks with the same height around $\bar{\Delta}=-\omega_{\mathrm{m}}$, in which the maximum value is less than those of the A-P and S-C cases. It indicates that the mode coupling increases the cooling limit at $\bar{\Delta}=-\omega_{\mathrm{m}}$ for the P-P case. This finding is completely different from the A-P case where the mode coupling greatly enhances the optical damping rate, corresponding to the great reduction cooling limit. We also find that a larger optomechanical coupling strength corresponds to a larger frequency shift $\delta_{\omega_{\mathrm{m}}}$ and optical damping $\Gamma_{\mathrm{opt}}$ for all three cases.

\section{Ground-state cooling without pre-cooling}

We now study the cooling limit of the mechanical resonator at the steady sate. The average phonon number with the Fock state probabilities $P_{n}$ is given as $\bar{n}=\sum_{0}^{\infty }nP_{n}$, where $n$ is the Fock state phonon number. Solving the rate equation in the steady state (see Appendix~\ref{appB}) with $\dot{\bar{n}}=0$, the final phonon number $n_{\mathrm{f}}$ will be
\begin{equation}
n_{\mathrm{f}}=\frac{\gamma_{\mathrm{m}}n_{\mathrm{th}}+A_{+}}{\gamma_{\mathrm{m}}+\Gamma_{\mathrm{opt}}}, \label{Nfinal}
\end{equation}
which can be divided into two parts: (i) the classical cooling limit
$n_{\mathrm{f}}^{\mathrm{c}}=\gamma_{\mathrm{m}}n_{\mathrm{th}}/(\gamma_{\mathrm{m}}+\Gamma_{\mathrm{opt}})$; and (ii) the quantum cooling limit $n_{\mathrm{f}}^{\mathrm{q}}=A_{+}/(\gamma_{\mathrm{m}}+\Gamma_{\mathrm{opt}})$.

\begin{figure}[ptb]
\includegraphics[bb=30 200 531 625,  width=9cm, clip]{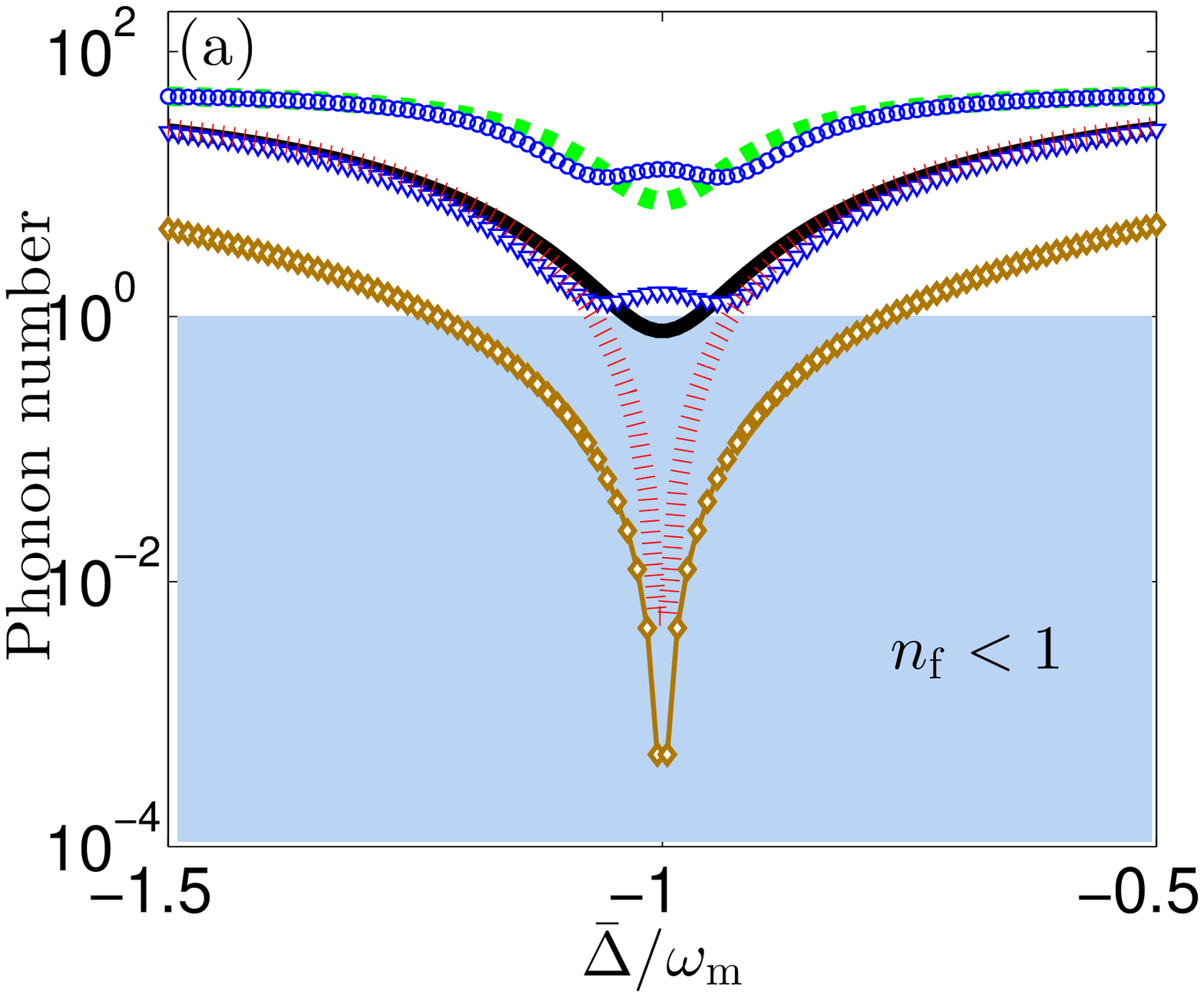}
\includegraphics[bb=30 200 531 625,  width=9cm, clip]{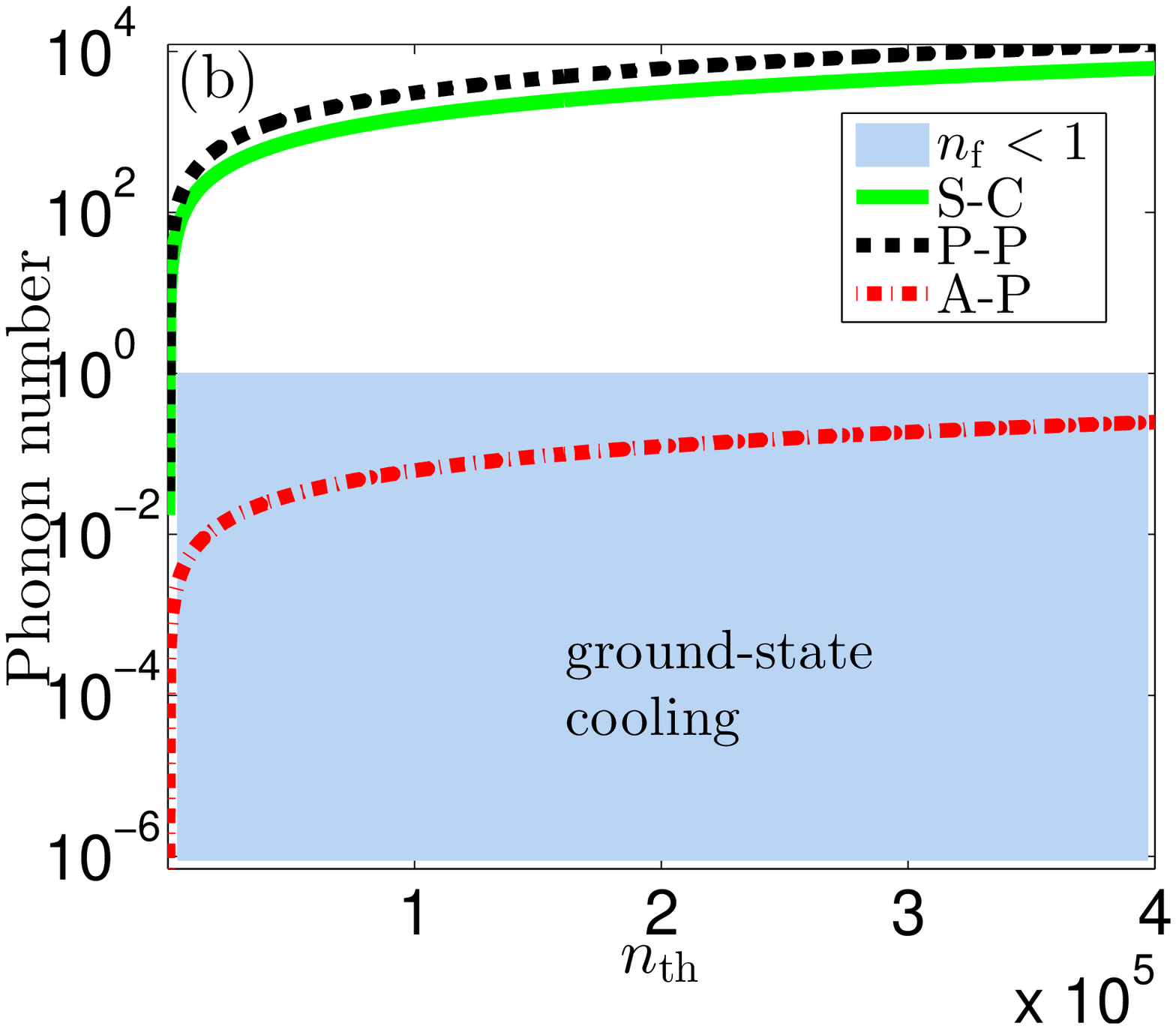}
\caption{(Color online) (a) The phonon number $n_{\mathrm{f}}$ versus $\bar{\Delta}$ with thermal photon number $n_{\mathrm{th}}=50$ for three cases (A-P, S-C, and P-P). Two coupling strengths $G_{\mathrm{lin}}/\gamma=0.015$ and $0.05$ are considered and represented by green-dashed and black-solid curves for the S-C case, blue-circle  and blue-triangle curves for the P-P case, red-dotted  and brown-rhombus curves for the A-P case, respectively. (b) The minimum phonon number $n_{\mathrm{f,\min}}$ versus $n_{\mathrm{th}}$ with $G_{\mathrm{lin}}/\gamma=$ $0.05$, and $\bar{\Delta}=-\omega_{\mathrm{m}}$. Other parameters are the same as those in Fig.~\ref{fig2}.}
\label{fig6}
\end{figure}

As shown in Fig.~\ref{fig6}(a), the optimal cooling limit with minimum $n_{\mathrm{f}}$ locates at the red detuning point $\bar{\Delta}=-\omega_{\mathrm{m}}$ for the S-C case. When $G_{\mathrm{lin}}/\gamma=0.05$, the final cooling limit is $n_{\mathrm{f,\min}}=0.78$, which means the ground state cooling. However, when the optomechanical coupling strength is reduced to $G_{\mathrm{lin}}/\gamma=0.015$, then $n_{\mathrm{f,\min}}=8$, the ground state cooling cannot be achieved. This means, for the S-C case, the power of the control field should be strong enough so that the linearized optomechanical coupling $G_{\mathrm{lin}}$ exceeds the threshold $G_{\mathrm{lin}}^{\mathrm{th}}=\sqrt{\gamma\gamma_{\mathrm{m}}n_{\mathrm{th}}}/2$ for the ground state cooling. Here, $G_{\mathrm{lin}}^{\mathrm{th}}/\gamma$ $\simeq0.045$ for the S-C case with the parameters in Fig.~\ref{fig6}. For the P-P case, as shown in blue curves in Fig.~\ref{fig6}(a), the position of the optimal cooling shifts and splits into two dips. Compared to the S-C case, the final cooling limit increases and ground state cannot be achieved with both two coupling strengths $G_{\mathrm{lin}}/\gamma=0.015$ and $0.05$. Thus, the mode coupling is harmful to the ground state cooling in the P-P case. In contrast, for A-P case, the mode coupling greatly reduces the final cooling limit by four orders of magnitude. This is totally different from the P-P case, where the mode coupling increases the final cooling limit. Also, for both S-C and P-P cases,  the ground state cooling cannot be achieved with small optomechanical coupling strength, e.g., $G_{\mathrm{lin}}/\gamma$ $=0.015$. In contrast, for A-P system, the ground state cooling can still be achieved even for small optomechanical coupling strength, e.g., $G_{\mathrm{lin}}/\gamma=0.015$, corresponding to a lower pump power.

The cooling limit $n_{\mathrm{f}}$ is mostly determined by the classical cooling limit $n_{\mathrm{f}}^{\mathrm{c}}=\gamma_{\mathrm{m}}n_{\mathrm{th}}/\left(\gamma_{\mathrm{m}}+\Gamma_{\mathrm{opt}}\right)  $ for all S-C, P-P, and A-P cases. As shown in Fig.~\ref{fig5}(f), the optical damping $\Gamma_{\mathrm{opt}}$ of the mechanical resonator is enhanced by four orders of magnitude for A-P case. Thus, the classical cooling limit is significantly lowered. This means that the the condition for initially cryogenic pre-cooling and quality factor of the mechanical resonator can be greatly relaxed, i.e., a higher bath thermal phonon number $n_{\mathrm{th}}$ and large mechanical decay rate $\gamma_{\mathrm{m}}$ can be tolerated. The ground-state cooling is marked by blue-shaded region in Fig.~\ref{fig6}(b). The optimal cooling limit (or saying minimum phonon number) $n_{\mathrm{f,\min}}$ quickly becomes larger than one when the thermal phonon number $n_{\mathrm{th}}$ is increased for S-C and P-P cases. For S-C and P-P cases, the initial upper bound thermal phonon number $n_{\mathrm{th}}^{\mathrm{t}}$, for achieving final optimal phonon number $n_{\mathrm{f,\min}}=1$ via the cooling, is given by $n_{\mathrm{th}}^{\mathrm{t}}=65$ and $32$, respectively. The corresponding bath temperatures are $T_{\mathrm{c}}=31.2$ mK and $63$ $\mathrm{mK}$ when the frequency of the mechanical resonator is $\omega_{\mathrm{m}}/2\pi=20$ $\mathrm{MHz}$. However, for the A-P case, the final optimal cooling limit $n_{\mathrm{f,\min}}$ is still less than one even with a very large initially thermal phonon number, e.g., $n_{\mathrm{th}}=305260$, corresponding to $T=293$ K (i.e., the room temperature). Thus the requirement for initially thermal phonon number is greatly relaxed for the A-P case. This supports a pre-cooling free ground state cooling of mechanical resonator.

\section{Conclusion}

We propose a method to cool the mechanical resonator of the optomechanical system to its ground state by coupling a gain cavity to the loss cavity of the optomechanical system. The coupled loss and gain cavities form into a $\mathcal{PT}$-symmetry system and result in optical field localization in the gain cavity (or saying active cavity).  Using the field localization mechanism, the heat of the mechanical resonator can be piped to the active cavity through the optomechanical coupling to the loss cavity, and the mechanical resonator can finally be cooled. We find that the optimal cooling rate is closely related to the phase transition of the $\mathcal{PT}$-symmetry system. Around the phase transition point, the net optical damping is enhanced at least four orders of magnitude, compared to that of standard optomechanical systems (the S-C case). Our cooling mechanism is also completely different from that of the standard optomechanical system coupled to an additional loss cavity (P-P case), where the mode coupling between two loss optical cavities (saying passive cavities) reduces the net optical damping and the cooling rate.

For our proposal, the energy localization induced giant optical damping for the mechanical resonator can be further applied to greatly reduce the coupling strength threshold or the minimum pump power of the control field for achieving the ground-state cooling. This  is totally different from the P-P cooling case, where both the threshold and the minimum pump power of the control field for achieving the ground-state cooling are increased by the mode coupling between two passive optical modes. Thus, our proposal can support a ultra-low power ground-state cooling of the mechanical resonator.

Using our proposal, the cooling limit can be greatly reduced around the phase transition point, compared to those of the S-C and the P-P cases. This in turn implies that the requirement for initially thermal phonon number can be greatly relaxed in our proposal. In particular, our proposal allows a ground-state cooling of the mechanical resonator without the need of the cryogenic precooling, e.g., allowing a direct cooling to the ground state from room temperature, i.e., 293 \textrm{K}, corresponding to very large thermal phonon number.

We finally mention that our active-cavity assisted optomechanical cooling scheme can also be extended to the hybrid optomechanical system consisting of two-level atomic ensemble inside the optical cavity. In this case, the two-level atomic ensemble plays a role of the gain cavity.

In summary, our proposal has the potential to allow an ultra-low-power and pre-cooling-free cooling of the mechanical resonator towards its quantum ground state, and provides a new way for quantum mechanical manipulation of macroscopic mechanical resonators at the room temperature.


\section{Acknowledgement}
Y.X.L. is supported by the National Basic Research Program of China 973 Program under Grant No.~2014CB921401, the Tsinghua University Initiative Scientific Research Program, and the Tsinghua National Laboratory for Information Science and Technology (TNList) Cross-discipline Foundation.

\appendix
\section{Linearization and effective Hamiltonian}\label{appa}

We study the system of a mechanical resonator and two coupled optical cavities, in which one has loss and another one has gain. The mechanical resonator is optomechanically coupled to the lossy (or saying passive) optical cavity, which is coherently driven by an external laser filed. This optomechanically coupled mechanical resonator and cavity field system forms a standard optomechanical system. The coherent pump of the passive cavity mode is described by $H_{\mathrm{pump}}=\Omega e^{-i\omega_{\mathrm{p}}t}b^{\dag}+\Omega^{\ast}e^{i\omega_{\mathrm{p}}t}b$, where $\Omega$ denotes the driving strength. Using displacement transformations $a\rightarrow\alpha_{1}+a,b\rightarrow\alpha_{2}+b,c\rightarrow\beta+c$, the optomechanical coupling between mode $b$ and mechanical mode $c$ can be linearized and the interaction Hamiltonian reads
\begin{equation}
H_{\mathrm{opt}}=\hbar G_{\mathrm{lin}}(b^{\dag}+b)(c+c^{\dag}), \label{Hopt}%
\end{equation}
where $G_{\mathrm{lin}}\equiv g\alpha_{2}$. In the rotating reference frame at the driving field frequency $\omega_{\mathrm{p}}$ and with linearization approach, the Hamiltonian of the whole system in Eq.~(\ref{SPhoton}) is then written as an effective Hamiltonian
\begin{align}
H_{\mathrm{lin}}  &  =-\hbar\Delta_{\mathrm{a}}a^{\dag}a-\hbar\Delta_{\mathrm{b}}^{^{\prime}}b^{\dag}b+\hbar J(a^{\dag}b+b^{\dag}a)\nonumber\\
&  +\hbar\omega_{\mathrm{m}}c^{\dag}c+\hbar\left(  G_{\mathrm{lin}}^{\ast}b+G_{\mathrm{lin}}b^{\dag}\right)  \left(  c+c^{\dag}\right)  , \label{Hlin}
\end{align}
with the modified detunings
\begin{align}
\Delta_{\mathrm{a}}  &  =\omega_{\mathrm{p}}-\omega_{\mathrm{a}},\\
\Delta_{\mathrm{b}}  &  =\omega_{\mathrm{p}}-\omega_{\mathrm{b}},\\
\Delta_{\mathrm{b}}^{^{\prime}}  &  =\Delta_{\mathrm{b}}-g\left(  \beta
+\beta^{\ast}\right)  .
\end{align}
Using equations of motion in Eqs.~(3-5), we can obtain steady state values $\alpha_{1}$, $\alpha_{2}$, and $\beta$ of the
variables $a$, $b$, and $c$ as
\begin{align}
\alpha_{1}  &  =-i\frac{J\alpha_{2}}{i\Delta_{\mathrm{a}}-\kappa/2}
,\\
\alpha_{2}  &  =\frac{\Omega}{\gamma/2+i\Delta_{\mathrm{b}}^{^{\prime}}+\lambda},\\
\beta &  =-i\frac{g\left\vert \alpha_{2}\right\vert ^{2}}{\gamma_{\mathrm{m}%
}/2+i\omega_{\mathrm{m}}}.
\end{align}
with $\lambda=J^{2}/\left(  i\Delta_{\mathrm{a}}-\kappa/2\right)  $. Here, $\kappa$ is the gain of mode $a$, and $\gamma$ and $\gamma_{\mathrm{m}}$ are the decay rates of the modes $b$ and $c$, respectively. Without loss of generality, we assume  $\Delta_{\mathrm{a}}=\Delta_{\mathrm{b}}^{^{\prime}}=\bar{\Delta}$.

\section{Optical force spectrum}\label{appb}

With the definition of fourier transformation
\begin{align}
f\left(  \omega\right)&  =\frac{1}{\sqrt{2\pi}}\int_{-\infty}^{+\infty}f\left(  t\right)  e^{i\omega t}dt, \label{appb1}\\
f\left(  t\right)&  =\frac{1}{\sqrt{2\pi}}\int_{-\infty}^{+\infty}f\left(\omega\right) e^{-i\omega t}d\omega, \label{appb2}
\end{align}
we obtain noise operators $a_{\mathrm{in}}\left(\omega\right)$, $b_{\mathrm{in}}\left(\omega\right)$, and $c_{\mathrm{in}}\left(  \omega\right)$ in the frequency domain as follows
\begin{align}
a_{\mathrm{in}}\left(  \omega\right)   &  =\frac{1}{\sqrt{2\pi}}\int_{-\infty
}^{+\infty}a_{\mathrm{in}}\left(  t\right)  e^{i\omega t}dt, \label{appb3} \\
b_{\mathrm{in}}\left(  \omega\right)   &  =\frac{1}{\sqrt{2\pi}}\int_{-\infty
}^{+\infty}a_{\mathrm{in}}\left(  t\right)  e^{i\omega t}dt, \label{appb4} \\
c_{\mathrm{in}}\left(  \omega\right)   &  =\frac{1}{\sqrt{2\pi}}\int_{-\infty
}^{+\infty}c_{\mathrm{in}}\left(  t\right)  e^{i\omega t}dt. \label{appb5}
\end{align}
The correlation function for noise operator $a_{\mathrm{in}}(\omega)$ in the frequency domain is given by
\begin{align}
& \left\langle a_{\mathrm{in}}^{\dag}\left(  \omega\right)  a_{\mathrm{in}}\left(  \omega^{^{\prime}}\right)  \right\rangle \\
& =\frac{1}{2\pi}\int_{-\infty}^{+\infty}\left\langle a_{\mathrm{in}}^{\dag}\left(  t\right)  a_{\mathrm{in}}\left(  t^{^{\prime}}\right)  \right\rangle e^{i\omega t}dt\int_{-\infty}^{+\infty}e^{i\omega^{^{\prime}}t^{^{\prime}}}dt^{^{\prime}}, \label{appb6}\\
& \left\langle a_{\mathrm{in}}\left(  \omega\right)  a_{\mathrm{in}}^{\dag
}\left(  \omega^{^{\prime}}\right)  \right\rangle \\
& =\frac{1}{2\pi}\int_{-\infty}^{+\infty}\left\langle a_{\mathrm{in}}\left(t\right)  a_{\mathrm{in}}^{\dag}\left(  t^{^{\prime}}\right)  \right\rangle
e^{i\omega t}dt\int_{-\infty}^{+\infty}e^{i\omega^{^{\prime}}t^{^{\prime}}
}dt^{^{\prime}}.\label{appb7}
\end{align}
Using correlation function in Eqs.~(\ref{ano1})-(\ref{ano2}) in time domain, we can further obtain
\begin{align}
\left\langle a_{\mathrm{in}}^{\dag}\left(  \omega\right)  a_{\mathrm{in}}\left(  \omega^{^{\prime}}\right)  \right\rangle  & =\delta\left(\omega+\omega^{^{\prime}}\right), \label{appb8}\\
\left\langle a_{\mathrm{in}}\left(  \omega\right)  a_{\mathrm{in}}^{\dag}\left(  \omega^{^{\prime}}\right)  \right\rangle  & =0. \label{appb9}
\end{align}
Following the same procedures, the correlation functions of the noise operators $b_{\mathrm{in}}$ and $c_{\mathrm{in}}$ in the frequency domain are given as
\begin{align}
\left\langle b_{\mathrm{in}}^{\dag}\left(  \omega\right)  b_{\mathrm{in}}\left(  \omega^{^{\prime}}\right)  \right\rangle  & =0,\label{appb10}\\
\left\langle b_{\mathrm{in}}\left(  \omega\right)  b_{\mathrm{in}}^{\dag}\left(  \omega^{^{\prime}}\right)  \right\rangle  & =\delta\left(\omega+\omega^{^{\prime}}\right)  ,\label{appb11}\\
\left\langle c_{\mathrm{in}}^{\dag}\left(  \omega\right) c_{\mathrm{in}}\left(  \omega^{^{\prime}}\right)  \right\rangle  & =n_{\mathrm{th}}(\delta\left(  \omega+\omega^{^{\prime}}\right)  ),\label{appb12}\\
\left\langle c_{\mathrm{in}}\left(  \omega\right) c_{\mathrm{in}}^{\dag}\left(  \omega^{^{\prime}}\right)  \right\rangle  & =(n_{\mathrm{th}}+1)\delta\left(  \omega+\omega^{^{\prime}}\right)  .\label{appb13}
\end{align}
Compared to Eqs.~(\ref{appb8})-(\ref{appb9}), it is notable that the noise correlation functions of active mode $a$ are totally different from that of the passive mode $b$. The optical force acting on the mechanical resonator is descried by
\begin{equation}
F(t)=-\frac{\hbar}{x_{\mathrm{ZPF}}}[G_{\mathrm{lin}}^{\ast}b(t)+G_{\mathrm{lin}}b^{\dag}(t)],
\end{equation}
and the quantum noise spectrum is calculated by
\begin{equation}
S_{FF}\left( \omega \right) =\int_{-\infty }^{+\infty }d\tau e^{i\omega \tau}\left\langle F(t)F(t^{^{\prime }})\right\rangle ,  \label{appb14}
\end{equation}
where $\tau =t-t^{^{\prime }}$. Using the following transformation
\begin{align}
F(t)  & =\frac{1}{\sqrt{2\pi}}\int_{-\infty}^{+\infty}F\left(  \omega\right)e^{-i\omega t}d\omega,\label{appb15}\\
F(t^{^{\prime}})  & =\frac{1}{\sqrt{2\pi}}\int_{-\infty}^{+\infty}F\left(\omega^{^{\prime}}\right)  e^{-i\omega^{^{\prime}}t^{^{\prime}}}d\omega^{^{\prime}},\label{appb16}
\end{align}
we can obtain the quantum noise spectrum of optical force in the frequency domain as
\begin{align}
& S_{FF}\left( \omega \right)   \notag \\
& =\frac{1}{2\pi }\int_{-\infty }^{+\infty }d\tau e^{i\omega \tau}\int_{-\infty }^{+\infty }\int_{-\infty }^{+\infty }\left\langle F\left(
\omega \right) F\left( \omega ^{^{\prime }}\right) \right\rangle   \notag \\
& \times e^{-i\omega ^{^{\prime }}t^{^{\prime }}}e^{-i\omega t}d\omega^{^{\prime }}d\omega ,  \label{appb17}
\end{align}
where
\begin{equation}
F(\omega)=-\frac{\hbar}{x_{\mathrm{ZPF}}}\left[  G_{\mathrm{lin}}^{\ast}b(\omega)+G_{\mathrm{lin}}b^{\dag}(\omega)\right].
\label{appb177}
\end{equation}
By solving Eqs.~(\ref{Lw1})-(\ref{Lw3}), we can further obtain the correlation functions of $F\left(\omega\right)$, which is given as
\begin{equation}
\left\langle F\left(  \omega\right) F\left(\omega^{^{\prime}}\right)\right\rangle =X_{FF}(\omega,\omega^{^{\prime}})\delta\left(\omega+\omega^{^{\prime}}\right),\label{appb18}
\end{equation}
with
\begin{eqnarray}
X_{FF}(\omega,\omega^{^{\prime}})=\frac{\hbar^2\left\vert G_{\mathrm{lin}}\right\vert ^{2}}{x_{\mathrm{ZPF}}^{2}}[\gamma \chi(\omega,-\omega^{\prime})+J^{2}\kappa \Upsilon(-\omega,\omega^{\prime})],\nonumber\\
\label{appb19}
\end{eqnarray}
with $\chi(\omega,-\omega^{\prime})=\chi\left(  \omega\right)  \chi^{\ast}(  -\omega^{^{\prime}%
})$ and $\Upsilon(-\omega,\omega^{\prime})=\chi^*\left( -\omega\right)  \chi( \omega^{^{\prime}%
})\chi^*_{a}\left( -\omega\right)  \chi_{a}( \omega^{^{\prime}})$. By taking Eq.~(\ref{appb18}) and (\ref{appb19}) into Eq.~(\ref{appb17}), we have
\begin{align}
& S_{FF}\left(  \omega\right)  \nonumber\\
& =\frac{1}{2\pi}\int_{-\infty}^{+\infty}e^{i\omega\tau}d\tau\int_{-\infty
}^{+\infty}X_{FF}(\omega,\omega^{^{\prime}}\rightarrow-\omega)e^{-i\omega\tau
}d\omega\nonumber\\
& =X_{FF}(\omega,\omega^{^{\prime}}\rightarrow-\omega).\label{appb24}
\end{align}
Combing Eq.~(\ref{appb19}) and Eq.~(\ref{appb24}), we can obtain the quantum noise spectrum of the optical force as given in Eq.~(\ref{SFFW}) of the main text.

\section{Cooling and heating rates}\label{appB}

The photon number fluctuations give rise to noisy force on the mechanical resonator via the optomechanical interaction, which will induce transitions between energy eigenstates (phonon Fock states) of the mechanical resonator. Here, the linearized
optomechanical coupling strength $G_{\mathrm{lin}}$ can be controlled by changing the driving strength of the control field. If this coupling strength is small enough, then the noise effect can be treated by using the perturbation theory, and the transition rates can be obtained by virtue of Fermi's Golden Rule. We assume that the state of the mechanical resonator is $\left\vert \Psi (t)\right\rangle $. In the interaction picture and to the first-order of the interaction between the cavity field and the mechanical resonator, we have
\begin{equation}
\left\vert \Psi _{I}(t)\right\rangle =\left\vert \Psi _{I}(0)\right\rangle -\frac{i}{\hbar }\int_{0}^{t}d\tau \hat{V}(\tau )\left\vert \Psi
_{I}(0)\right\rangle ,  \label{INT}
\end{equation}
where $\hat{V}(\tau )=\hbar \left( G_{\mathrm{lin}}^{\ast }b+G_{\mathrm{lin}}b^{\dag }\right) \left( c+c^{\dag }\right) $ is the linearized optomechanical coupling. If the mechanical resonator is initially in $n$ phonon state, the amplitude for transition to $n+1$ phonon state (i.e.,generating an extra phonon) at the time $t$ can be obtained from Eq.~(\ref{INT}), and is given as
\begin{equation}
\alpha _{n\rightarrow n+1}=-i\int_{0}^{t}d\tau \left( G_{\mathrm{lin}}^{\ast}b+G_{\mathrm{lin}}b^{\dag }\right) \left\langle n+1\right\vert \left(c+c^{\dag }\right) \left\vert n\right\rangle.  \label{amp1}
\end{equation}
Using the optical force operator $F(t)$ defined in Eq.~(\ref{Ll1}), we can further express Eq.~(\ref{amp1}) as
\begin{equation}
\alpha _{n\rightarrow n+1}=\frac{i}{\hbar }x_{\mathrm{ZPF}}\sqrt{n+1}%
\int_{0}^{t}e^{i\omega _{\mathrm{m}}\tau }d\tau F(\tau ).  \label{amp2}
\end{equation}
Thus, the probability for the transition of the mechanical resonator from $n$ phonon state to $n+1$ phonon state through the Stokes scattering is
\begin{eqnarray}
p_{n\rightarrow n+1} &=&\left\vert \alpha _{n\rightarrow n+1}\right\vert
^{2}=\frac{x_{\mathrm{ZPF}}^{2}}{\hbar ^{2}}\left( n+1\right)    \\
&&\times \int_{0}^{t}\int_{0}^{t}e^{-i\omega _{\mathrm{m}}\left( \tau
_{1}-\tau _{2}\right) }d\tau _{1}d\tau _{2}F(\tau _{1})F(\tau _{2}).\notag
\label{Prob1}
\end{eqnarray}
By taking ensemble average for the cavity field, we have
\begin{eqnarray}
P_{n\rightarrow n+1} &=&\frac{x_{\mathrm{ZPF}}^{2}}{\hbar ^{2}}\left(
n+1\right)   \\
&&\times \int_{0}^{t}\int_{0}^{t}e^{-i\omega _{\mathrm{m}}\left( \tau
_{1}-\tau _{2}\right) }d\tau _{1}d\tau _{2}\left\langle F(\tau _{1})F(\tau
_{2})\right\rangle .\label{Probn} \notag
\end{eqnarray}
This calculation is closely related to that used to discuss the transition from the ground state to the excited state for an atomic spectrum analyzer~\cite{speca} or cavity assisted laser cooling of trapped ions~\cite{atomcooling1}, atomic and molecular motions systems~\cite{atomcooling2}.

We consider that the integration time $t$ is much longer than the autocorrelation time $\tau _{c}$ (i.e., $t\gg \tau _{c}$).
Using the Fourier transform defined in Eq.~(\ref{appb1}) and the Wiener-Khinchin theorem, we find that Eq.~(\ref{Probn}) can be further given as
\begin{equation}
P_{n\rightarrow n+1}(t)=\frac{x_{\mathrm{ZPF}}^{2}}{\hbar ^{2}}\left(n+1\right) S_{FF}\left( -\omega _{\mathrm{m}}\right) t,  \label{Probf}
\end{equation}
where
\begin{equation}
S_{FF}\left( \omega \right) =\int_{-\infty }^{+\infty }d\tilde{\tau}%
e^{i\omega \tilde{\tau}}\left\langle F(\tau _{1})F(\tau _{2})\right\rangle .
\label{sprm}
\end{equation}
is the quantum noise spectrum of the optical force, as shown in Eq.~(\ref{appb14}). Here, $\tilde{\tau}=\tau _{1}-\tau _{2}$ is the time interval. We can find that the probability $P_{n\rightarrow n+1}$ indeed increases linearly with time under the long
time assumptions.

Following the same procedure, we can obtain the rate for optically induced transitions from $n+1$ phonon state to $n$ phonon state through the anti-Stokes scattering (absorbing a phonon from the mechanical resonator in the process), and the probability is given as
\begin{equation}
P_{n+1\rightarrow n}(t)=\frac{x_{\mathrm{ZPF}}^{2}}{\hbar ^{2}}\left(n+1\right) S_{FF}\left( \omega _{\mathrm{m}}\right) t.  \label{gm1}
\end{equation}
The time derivative of the probability gives the transition rates
\begin{eqnarray}
\Gamma _{n\rightarrow n+1} &=&\left( n+1\right) \frac{x_{\mathrm{ZPF}}^{2}}{\hbar ^{2}}S_{FF}\left( -\omega _{\mathrm{m}}\right) , \\
\Gamma _{n+1\rightarrow n} &=&\left( n+1\right) \frac{x_{\mathrm{ZPF}}^{2}}{\hbar ^{2}}S_{FF}\left( \omega _{\mathrm{m}}\right) ,
\end{eqnarray}
corresponding to the rate of generating and absorbing one phonon, respectively. We can also chose to follow the notation%
\begin{eqnarray}
A_{+} &=&\frac{x_{\mathrm{ZPF}}^{2}}{\hbar ^{2}}S_{FF}\left( -\omega _{\mathrm{m}}\right) , \\
A_{-} &=&\frac{x_{\mathrm{ZPF}}^{2}}{\hbar ^{2}}S_{FF}\left( \omega _{\mathrm{m}}\right) ,
\end{eqnarray}
which are widely used in the literatures. Given the rates $A_{+}$ ($A_{-}$) for generating  (annihilating) phonons in the
mechanical resonator, the net optical damping for the mechanical resonator is
\begin{equation}
\Gamma _{\mathrm{opt}}=A_{-}-A_{+}.  \label{gmanet}
\end{equation}

According to these transition rates, we may write the rate equation for the probability $P_{n}(t)$  of the Fock state that there are $n$ quanta in the mechanical resonator,
\begin{eqnarray}
\dot{P}_{n} &=&\Gamma _{n+1\rightarrow n}P_{n+1}+\Gamma _{n-1\rightarrow
n}P_{n-1}  \notag \\
&&-\Gamma _{n\rightarrow n-1}P_{n}-\Gamma _{n\rightarrow n+1}P_{n}  \notag \\
&&+\gamma _{\mathrm{m}}\left( n_{\mathrm{th}}+1\right) (n+1)P_{n+1}+\gamma _{%
\mathrm{m}}n_{\mathrm{th}}nP_{n-1}  \notag \\
&&-\gamma _{\mathrm{m}}\left( n_{\mathrm{th}}+1\right) nP_{n}-\gamma _{%
\mathrm{m}}n_{\mathrm{th}}\left( n+1\right) P_{n}.  \label{RateE}
\end{eqnarray}

The average phonon number is given as $\bar{n}=\sum_{n=0}^{\infty }nP_{n}$. Combined with the rate equation (\ref{RateE}), the infinite set of rate equations can be replaced by a single differential equation for the generating function $F(z,t)=P_{n}(t)z^{n}$, leading to
\begin{equation}
\dot{\bar{n}}=(\bar{n}+1)(A_{+}+A_{+}^{\mathrm{th}})-\bar{n}(A_{-}+A_{-}^{%
\mathrm{th}}),
\end{equation}
where $A_{+}^{\mathrm{th}}=n_{\mathrm{th}}\gamma _{\mathrm{m}}$, and $A_{-}^{\mathrm{th}}=\left(n_{\mathrm{th}}+1\right)\gamma _{\mathrm{m}}$ are the extra transitions rates due to the mechanical resonator's thermal environment, which has a mean thermal phonon number $n_{\mathrm{th}}$. The final phonon number $n_{\mathrm{f}}$ in the steady requiring $\dot{\bar{n}}=0$, will be
\begin{equation}
n_{\mathrm{f}}=\frac{A_{+}+n_{\mathrm{th}}\gamma _{\mathrm{m}}}{\Gamma _{\mathrm{opt}}+\gamma _{\mathrm{m}}}.
\end{equation}
Note that we have verified that the optomechanical coupling strength $G_{\mathrm{lin}}$ used in the main text is small enough for a perturbation treatment. Furthermore, the optomechanical interaction and the mechanical effect for the quantum noise spectrum of the optical force can also be neglected with the optomechanical coupling strength parameters used in the main text.

\section{Active-cavity assisted optical spring effect}\label{appc}
By solving Eqs.~(\ref{Lw1})-(\ref{Lw3}), we can also obtain the solution of $c(\omega)$ as
\begin{equation}
c\left(  \omega\right)  \simeq\frac{\sqrt{\gamma_{\mathrm{m}}}c_{\mathrm{in}}\left(  \omega\right)  -i\sqrt{\gamma}A\left(  \omega\right)  -\sqrt{\kappa}B\left(  \omega\right)  }{i\omega-i\left[  \omega_{\mathrm{m}}+\sum\left(\omega\right)  \right]  -\gamma_{\mathrm{m}}/2}, \label{COOL1}
\end{equation}
where
\begin{align}
A\left(  \omega\right)   &  =G_{\mathrm{lin}}^{\ast}\chi\left(  \omega\right) b_{\mathrm{in}}\left(  \omega\right)  +G_{\mathrm{lin}}\chi^{\ast}\left(-\omega\right)  b_{\mathrm{in}}^{\dag}\left(  \omega\right)  ,\label{COOL2}\\
B\left(  \omega\right)   &  =JG_{\mathrm{lin}}^{\ast}\chi\left(\omega\right)  \chi_{\mathrm{a}}\left(  \omega\right)  a_{\mathrm{in}}\left(\omega\right) \nonumber \\
& -JG_{\mathrm{lin}}\chi^{\ast}\left(  -\omega\right)\chi_{\mathrm{a}}^{\ast}\left(  -\omega\right)  a_{\mathrm{in}}^{\dag}\left(
\omega\right)  , \label{COOL3}
\end{align}
and
\begin{align}
\sum\left(  \omega\right)   &  =-i\left\vert G_{\mathrm{lin}}\right\vert^{2}\left[  \chi\left(  \omega\right)  -\chi^{\ast}\left(  -\omega\right)\right]  ,\label{COOL4}\\
\chi\left(  \omega\right)   &  =\left[  \chi_{\mathrm{b}}^{-1}\left(\omega\right)  +J^{2}\chi_{\mathrm{a}}\left(  \omega\right)  \right]^{-1}.\label{COOL5}
\end{align}
Here, $\sum\left(\omega\right)$ is the optomechanical self-energy, and $\chi\left(\omega\right)$ is the total response function of the coupled active and passive cavities.

The optomechanical coupling induced mechanical frequency shift and the net damping are given by
\begin{equation}
\delta_{\mathrm{\omega}_{\mathrm{m}}}=\operatorname{Re}\left[\sum(\omega_{\mathsf{m}})\right],
\end{equation}
and
\begin{equation}
\Gamma_{\mathrm{opt}}=-2\operatorname{Im}\left[\sum(\omega_{\mathsf{m}})\right].
\end{equation}

Using Eqs.~(\ref{COOL1})-(\ref{COOL5}), we can further express $\delta_{\mathrm{\omega}_{\mathrm{m}}}$ and $\Gamma_{\mathrm{opt}}$ as
\begin{align}
\delta_{\mathrm{\omega}_{\mathrm{m}}}  &  =\left\vert G_{\mathrm{lin}}\right\vert ^{2}\frac{\left(\bar{\Delta}+\omega\right)  \left[\kappa_{\mathrm{d}}\left(\kappa_{\mathrm{d}}-\kappa_{\mathrm{c}}\right)-4L_{-}\right]}{4L_{-}^{2}+\left(  \bar{\Delta}+\omega\right)^{2}\left(\kappa_{\mathrm{d}}-\kappa_{\mathrm{c}}\right)^{2}}\nonumber\\
&+\left\vert G_{\mathrm{lin}}\right\vert ^{2}\frac{\left(\bar{\Delta}-\omega\right)\left[\kappa_{\mathrm{d}}\left(  \kappa_{\mathrm{d}}-\kappa_{\mathrm{c}}\right)-4L_{+}\right]}{4L_{+}^{2}+\left(\bar{\Delta
}-\omega\right)^{2}\left(\kappa_{\mathrm{d}}-\kappa_{\mathrm{c}}\right)^{2}}, \label{dwm}
\end{align}
and
\begin{align}
\Gamma_{\mathrm{opt}}  &  =-2\left\vert G_{\mathrm{lin}}\right\vert ^{2}\frac{\kappa_{\mathrm{d}}L_{-}+\left(  \bar{\Delta}+\omega\right)^{2}\left(\kappa_{\mathrm{d}}-\kappa_{\mathrm{c}}\right)}{4L_{-}^{2}+\left(
\bar{\Delta}+\omega\right)^{2}\left(\kappa_{\mathrm{d}}-\kappa_{\mathrm{c}}\right)^{2}}\nonumber\\
&+2\left\vert G_{\mathrm{lin}}\right\vert ^{2}\frac{\kappa_{\mathrm{d}}L_{+}+\left(\bar{\Delta}-\omega\right)  ^{2}\left(  \kappa_{\mathrm{d}}-\kappa_{\mathrm{c}}\right)}{4L_{+}^{2}+\left(\bar{\Delta}-\omega\right)^{2}\left(  \kappa_{\mathrm{d}}-\kappa_{\mathrm{c}}\right)^{2}},
\label{dgamma}
\end{align}
with
\begin{equation}
L_{\mp}=g^{2}-\kappa_{\mathrm{c}}\kappa_{\mathrm{d}}/4-\bar{\Delta}^{2}-\omega^{2}\mp2\bar{\Delta}\omega.
\end{equation}

\end{document}